\documentclass[a4paper,11pt]{article}
\pdfoutput=1 % if your are submitting a pdflatex (i.e. if you have
\usepackage{jheppub} % for details on the use of the package, please
\usepackage[T1]{fontenc} % if needed
\usepackage[utf8]{inputenc}

\usepackage[normalem]{ulem}

\usepackage{graphicx}
\usepackage{color}
\usepackage{amssymb}
\usepackage{amsthm}
\usepackage{amsmath}

\newcommand{\pa}{\partial}                     % partial derivative

\newcommand{\dg}{{2}}                            % half-dimension
\newcommand{\scs}{C}

\newcommand{\procaA}{\ensuremath{P}}
\newcommand{\procaF}{\ensuremath{\mathcal{F}}}

\newcommand{\be}{\begin{equation}}             %:skip:
\newcommand{\ee}{\end{equation}}               %:skip:
\newcommand{\ba}{\begin{eqnarray}}
\newcommand{\ea}{\end{eqnarray}}

\title{Massive vector fields in Kerr--Newman and Kerr--Sen black hole spacetimes}

\author[a,b]{Ramiro Cayuso,}
\emailAdd{rcayuso@perimeterinstitute.ca}

\author[c]{Oscar J. C. Dias,}
\emailAdd{ojcd1r13@soton.ac.uk}

\author[a,b]{Finnian Gray,}
\emailAdd{fgray@perimeterinstitute.ca}

\author[a,b]{David Kubiz\v n\'ak,}
\emailAdd{dkubiznak@perimeterinstitute.ca}

\author[a,b]{Aoibheann Margalit,}
\emailAdd{amargalit@perimeterinstitute.ca}

\author[d,e]{Jorge E. Santos,}
\emailAdd{jss55@cam.ac.uk}

\author[a,b,f]{Renato Gomes Souza,}
\emailAdd{rsouza@perimeterinstitute.ca}

\author[a,b]{Leander Thiele}
\emailAdd{lthiele@perimeterinstitute.ca}

\affiliation[a]{Perimeter Institute, 31 Caroline Street North, Waterloo, ON, N2L 2Y5, Canada}

\affiliation[b]{Department of Physics and Astronomy, University of Waterloo,
	Waterloo, Ontario, Canada, N2L 3G1}

\affiliation[c]{STAG research centre and Mathematical Sciences, Highfield Campus, University of Southampton, Southampton SO17 1BJ, UK}

\affiliation[d]{DAMTP, Centre for Mathematical Sciences, Wilberforce Road, Cambridge, CB3 0WA, United Kingdom}

\affiliation[e]{Institute for Advanced Study, Princeton, NJ 08540, USA}

\affiliation[f]{ICTP South American Institute for Fundamental Research, IFT-UNESP, Sao Paulo, SP, Brazil, 01140-070}

\date{18 December 2019}            % version 1.00; arxiv version 1

\abstract{The superradiant instability modes of ultralight massive vector bosons are studied for weakly charged rotating black holes in Einstein--Maxwell gravity (the Kerr--Newman solution) and low-energy heterotic string theory (the Kerr--Sen black hole). We show that in both these cases, the corresponding massive vector (Proca) equations can be fully separated, exploiting the hidden symmetry present in these spacetimes. The resultant ordinary differential equations are solved numerically to find the most unstable modes of the Proca field in the two backgrounds and compared to the vacuum (Kerr black hole) case.
}

\begin{document}
\maketitle
\section{Introduction}
\label{sc:intro}
Ultralight bosons feature in many different extensions of the standard model, such as string theory~\cite{Goodsell:2009xc}, and provide compelling candidates for explaining dark matter~\cite{Essig:2013lka}. One particular model for these ultralight bosons is a massive spin-1 particle known as the {\em Proca field}. Considered first by Proca~\cite{Proca:1936} as a way to understand short-range nuclear forces in flat spacetime (see also \cite{Belinfante:1949,Rosen:1994}), the Proca equation is presently an integral part of the Standard Model where it is used for describing the massive spin-1 $Z$ and $W$ bosons, as well can be generalized to curved spacetime, e.g.~\cite{Seitz:1986sc}.

Although direct detection of dark matter proves to be very difficult, recently a new line of investigation has opened up with a flurry of papers considering the interplay of these ultralight bosons and {\em superradiance} from black holes~\cite{Arvanitaki:2009fg,Arvanitaki:2014wva,Baryakhtar:2017ngi,Cardoso:2018tly,Brax:2019npi,Siemonsen:2019ebd}. In particular, it has been shown that the instabilities from these superradiant modes can, in principle, be used to detect beyond Standard Model particles~\cite{Arvanitaki:2009fg} and put bounds on the potential masses of dark matter candidates,  e.g. ~\cite{Baryakhtar:2017ngi}. For example, in the LIGO/LISA era ultralight bosons and superradiance can leave signatures in the signals of detected gravitational waves~\cite{Arvanitaki:2014wva,Brito:2017zvb,Guerra:2019srj,Palomba:2019vxe}; see \cite{Siemonsen:2019ebd} for the most recent fully relativistic calculation of the gravitational wave signals. The first step to studying these instabilities is to consider test fields on a background spacetime.
Naturally, almost all the previous studies have focused on, likely the most astrophysically relevant, rotating black hole of Einstein's general relativity -- the Kerr black hole solution.

However, taking seriously the low energy limits of string theory leads to new kinds of black holes, and, if one is extending the Standard Model to include ultralight bosons it is a natural question to ask how the superradiant instabilities generated by these particles are modified by extensions to general relativity.Such extensions lead to black holes that typically carry extra fields and charges. In the astrophysical $D=4$ dimensions the {\em Kerr--Sen geometry}~\cite{Sen:1992ua} arises from the low energy limit of heterotic string theory. It represents a black hole with mass $M$ and $U(1)$ charge $Q$ and contains two extra background fields; the scalar dilaton $\Phi$ and the 3-form $H$. In the limit that these two fields vanish the spacetime reduces to the Kerr black hole. On the other hand, this spacetime should be compared to the {\em Kerr--Newman solution}~\cite{Newman:1965my} which is the unique stationary black hole solution to the Einstein equations with $U(1)$ charge (it can also be understood as a solution of $N=2$, $D=4$ supergravity). Both Kerr--Newman and Kerr--Sen black holes are stationary and axisymmetric spacetimes, possessing two Killing vectors which aid in understanding the behaviour of test fields in these backgrounds. In the Kerr--Newman case there exists an additional hidden symmetry of the {\em principal} Killing--Yano tensor which gives rise to Carter's constant for charged geodesics~\cite{FrolovKrtousKubiznak:2017review}. For the Kerr--Sen spacetime only a {\em generalized principal tensor} with torsion exists which is a weaker but still rather useful structure \cite{Houri:2010fr}.

The aim of this paper is to study the {\em superradiant instabilities} of the Kerr--Sen and Kerr--Newman black holes, as triggered by the ultralight massive bosons. These are well understood in the case of massive scalar fields, see \cite{Huang:2017whw} and \cite{Huang:2018qdl}, but the corresponding study for massive vectors is currently missing. The reason is simple. Even for vacuum (Kerr) black holes, the corresponding Proca equations are rather complicated partial differential equations whose direct decoupling and separation \`a la Teukolsky \cite{Teukolsky:1972,Teukolsky:1973} does not work due to the presence of the mass term~
\cite{Pani:2012vp,Rosa:2011my}. As a consequence the problem was investigated either using approximations~\cite{Baryakhtar:2017ngi,Pani:2012vp,Rosa:2011my} or employing serious numerical analysis~\cite{Cardoso:2018tly,East:2017ovw,East:2017mrj}.

However, a separability renaissance for vector fields has begun in the last couple of years due to a new ansatz by Lunin~\cite{Lunin:2017}. Simplified and written in covariant form by Frolov--Krtou\v{s}--Kubiz\v{n}\'ak~\cite{KrtousEtal:2018,FrolovKrtousKubiznak:2018a}, the new ansatz works for the massive vector field case~\cite{Frolov:2018ezx} and can be applied in the Kerr--NUT--AdS spacetimes for all dimensions. Importantly the Lunin--Frolov--Krtou\v{s}--Kubiz\v{n}\'ak (LFKK) ansatz exploits the existence of hidden symmetries in these spacetimes which are encoded in the principal tensor \cite{FrolovKrtousKubiznak:2017review} and allows the Proca equations to be decoupled and separated into ordinary differential equations \cite{Frolov:2018ezx}. In four dimensions, the separation equally  applies to the non-accelerating electro-vacuum type D Pleba{\' n}ski--Demia{\' n}ski spacetimes and thence to the Kerr--Newman black holes.\footnote{See \cite{Frolov:2018ezx, Dolan:2019hcw, Dolan:2018dqv} for further discussions of the LFKK ansatz in four dimensions, and \cite{Houri:2019nun} for an alternative way of separating the Maxwell equations in the Wahlquist metrics.}

More recently, the LFKK ansatz has been applied to study the Proca equations in the background of the Chong--Cveti{\v c}--L{\" u}--Pope black hole of $D=5$ minimal gauged supergravity~\cite{Chong:2005hr} where, as for Kerr--Sen, the principal tensor must be generalized to the case with torsion and is a weaker construction~\cite{Kubiznak:2009qi,Houri:2017tlk}. In this case, the torsion is naturally identified with the Maxwell 3-form present in the spacetime, and both the principal tensor and the Proca equations pick up a corresponding torsion modification. Despite these significant differences, the LFKK ansatz still applies and the (torsion) modified Proca equations decouple and separate \cite{Cayuso:2019vyh}.

In this work, we will apply the LFKK ansatz to separate the Proca equations in the Kerr--Sen black hole background. Similar to the black hole of minimal gauged supergravity, the corresponding Proca equations have to be modified by the presence of torsion, now naturally identified with the 3-form field $H$. To account for the dilaton field $\Phi$, we work in the string frame.  The corresponding unstable superradiant modes are studied numerically, and compared to those of the Kerr--Newman spacetime. We do so in an astrophysically viable situation where the black holes are fast spinning (close to extremal) and weakly charged \cite{Volonteri_2005, PhysRevD.100.023007}.

The paper is organized as follows. In Sec.~\ref{sec:Geom} we review and compare the Kerr--Newman and Kerr--Sen spacetimes, and discuss their symmetries and extremality. In  Sec.~\ref{sec:Proca} we discus how the Proca equation is modified by the presence of the dilaton $\Phi$ and the 3-form $H$, and present the separated equations (a full derivation of which may be found in App.~\ref{sec:Separation}). In Sec.~\ref{sc:NumericsResults} we numerically investigate the separated equations and visualize the behaviour of the most unstable superradiant modes. We conclude in Sec.~\ref{sc:conclusion}.

%%%%%%%%%%%%%%%%%%%%%%%%%%%%%%%%%%%%%%%%%%%%%%%%%%%%%%%%%%%%%%%%%%%
%%%%%%%%%%%%%%%%%%%%%%%%%%%%%%%%%%%%%%%%%%%%%%%%%%%%%%%%%%%%%%%%%%%%%%
\section{Kerr--Sen and Kerr--Newman black holes}\label{sec:Geom}

In this section we present the Kerr--Sen and the Kerr--Newman metrics, the background spacetimes in which we will study the instability modes of the Proca equation. The two metrics we consider both describe rotating and charged black hole spacetimes, however there are some key differences in the Kerr--Sen case due to modifications of general relativity coming from the low energy heterotic string theory effective action.

%%%%%%%%%%%%%%%%%%%%%%%%%%%%%%%%%%%%
\subsection{Kerr--Newman geometry}

The Kerr--Newman solution \cite{Newman:1965my} is the most general solution of the Einstein--Maxwell equations for an asymptotically flat, stationary and axisymmetric black hole. Its line element and vector potential read:

\begin{align} \label{metricKN}
\mathrm{d}s^2=&-\frac{\Delta}{\rho^2}\bigl(\mathrm{d}t-a\sin^2\!\theta \mathrm{d}\phi\bigr)^2+\frac{\rho^2}{\Delta}\mathrm{d}r^2 +\frac{\sin^2\!\theta}{\rho^2}\Bigl[a\,\mathrm{d}t-(r^2+a^2)\mathrm{d}\phi\Bigr]^2+\rho^2 \mathrm{d}\theta^2\,,\nonumber\\
A =&-\frac{Qr}{\rho^2}\bigl(\mathrm{d}t-a\sin^2\!\theta \mathrm{d}\phi\bigr)\,,
\end{align}
where
\be
\rho^2=r^2+a^2\cos^2\theta\,,\ \quad\   \Delta=r^2-2Mr+a^2 + Q^2\,.
\ee
The solution describes a black hole with mass $M$, charge $Q$, angular momentum  $J=Ma$, and a magnetic dipole moment $\mu_g = Qa$. The metric possesses a curvature singularity at $\rho^2=0$, which is protected by an event horizon at $r=r_{+}\equiv{}M + \sqrt{M^{2}-a^{2} - Q^{2}}$ provided that $a^{2} + Q^{2} \leq M^{2}$.  In the case where the equality holds we have an {\em extremal black hole}. The rotation of the black hole causes inertial frame dragging whose extreme manifestation is the existence of the ergosphere,
for $r_+<r<r_e\equiv M+\sqrt{M^2 -Q^2-a^2\cos^2\theta}$. In this region the time-like vector $\partial_t$ becomes null and therefore any massive particle must rotate. It is this region that leads to superradiant emission and is responsible for the instability modes for perturbations on this spacetime.

The black hole horizon rotates with angular velocity
\be\label{OmegaH}
\Omega_H=-\frac{g_{t\phi}}{g_{\phi\phi}}=\frac{a}{r_+^2+a^2}\;,
\ee
and can be assigned the following Hawking temperature, entropy, and electrostatic potential:
\be\label{TDsNewman}
T_H=\frac{\Delta'(r_+)}{r_+^2+a^2}=\frac{r_+^2-a^2-Q^2}{4\pi r_+(r_+^2+a^2)}\,,\quad
S=\pi(r_+^2+a^2)\,,\quad \phi_H=\frac{Qr_+}{r_+^2+a^2}\,.
\ee
These quantities satisfy the first law of black hole thermodynamics
\be\label{first}
\delta M=T_H\delta S+\Omega_H \delta J+\phi_H \delta Q\,,
\ee
as well as the associated Smarr relation, $M=2( T_H S+\Omega_H J)+\phi_H Q$\,.

The Kerr--Newman metric admits a hidden symmetry of the {\em principal tensor} (PT), which is a non-degenerate closed conformal Killing--Yano 2-form $h$,  obeying the following equation \cite{FrolovKrtousKubiznak:2017review}:
\be\label{PT}
\nabla_{c} h_{ab}=g_{ca}\xi_b-g_{cb}\xi_a\,,\quad \xi^a=\frac{1}{3}\nabla_c h^{ca}\,.
\ee
It is this tensor that underlies the separation of variables in the Proca equation. It explicitly reads as follows
\be\label{PTe}
h=r(\mathrm{d}t-a\sin^2\!\theta\,\mathrm{d}\phi)\wedge \mathrm{d}r-a\,\cos \theta \bigl[a\,\mathrm{d}t-(r^2+a^2)\mathrm{d}\phi\bigr]\wedge \mathrm{d}\cos\theta\,,
\ee
and gives rise to the associated Killing tensor
\be\label{KT}
K_{ab}=h_{ac}h^{c}{}_{b}-\frac{1}{2}g_{ab}h^2\;,
\ee
which generates the generalized Carter's constant for charged geodesics. It also yields the two independent isometries of the spacetime:
$\xi^a$ in \eqref{PT} and $\eta^a=K^{ab}\xi_b$ \cite{FrolovKrtousKubiznak:2017review}.

%%%%%%%%%%%%%%%%%%%%%%%%%%%%%%%%%
\subsection{Kerr--Sen geometry}

The Kerr--Sen black black hole \cite{Sen:1992ua} is an exact classical solution of the low-energy effective theory describing heterotic string theory given by the following action:

\begin{equation}\label{action}
S = \frac{1}{16\pi}\int \mathrm{d}x^{4}\,\sqrt{-g}\,e^{\Phi}\left(R + g^{a b}\partial_{a}\Phi\partial_{b}\Phi-F_{ab}F^{ab}-\frac{1}{12}H_{a b c}H^{a b c}\right)\,,
\end{equation}
where $g_{ab}$ represents the metric in the string frame, $\Phi$ is the dilaton field,  $F=\mathrm{d}A$ is the Maxwell field strength, and ${H} = \mathrm{d}\mathcal{B} -2A \wedge F$ is a 3-form defined in terms of the vector potential $A$ and a 2-form potential $\mathcal{B}.$\footnote{Note that we have rescaled the vector potential $A\to2\sqrt{2}A$ so that the Maxwell Lagrangian has the canonical prefactor~\cite{Sen:1992ua}.} The action is invariant under a $U(1)$ transformation $A\to A+\mathrm{d}\lambda$ provided we also send ${\cal B}\to {\cal B}+2\lambda F$ and  the corresponding equations of motion for the background fields $A$ and $H$ are,
\begin{equation}
\label{eq:HourieetalEq4.2}
\nabla^{a}\bigl[e^{\Phi}(F_{a b} - H_{a b c}A^{c})\bigr] = 0\,, \quad \nabla^a(e^{\Phi} H_{abc}) = 0\,.
\end{equation}
These will be important in section~\ref{sec:Proca} where we motivate a generalization of the Proca equation to this background. The full set of equations of motion is supplemented by the Einstein and dilaton  equations. Since these will not play any role in the further discussion we do not write them here explicitly and refer the interested reader to for example \cite{Houri:2010fr}.

In any case, the Kerr-Sen metric in the standard Boyer--Lindquist-type coordinates and the string frame reads \cite{Sen:1992ua,Wu:2001xh,Houri:2010fr}:
\begin{align}\label{Senmetric}
\mathrm{d}s^2\!&=\!e^{-\Phi}\Bigl\{-\frac{\Delta_b}{\rho_b^2}\bigl(\mathrm{d}t-a\sin^2\!\theta\,\mathrm{d}\phi\bigr)^2+\frac{\rho_b^2}{\Delta_b}\mathrm{d}r^2+\frac{\sin^2\!\theta}{\rho_b^2}\Bigl[a\,\mathrm{d}t-(r^2+2\,b\,r+a^2)\mathrm{d}\phi\Bigr]^2+\rho_b^2 \mathrm{d}\theta^2\Bigr\}\,,\nonumber\\
{\cal B}&=\frac{2\,b\,r}{\rho_b^2}\,a\,\sin^2\theta \mathrm{d}t\wedge \mathrm{d}\varphi\,,\quad
A\!=\!-\frac{Q\,r}{\rho^2}e^{-\Phi}\bigl(\mathrm{d}t-a\,\sin^2\!\theta \mathrm{d}\phi\bigr)\,,\quad
e^{-\Phi}= \frac{\rho^2}{\rho_b^2}\,,
\end{align}
where the metric functions are given by
\be
\rho^2=r^2+a^2\cos^2\theta\,,\quad   \rho_b^2=\rho^2+2br\,,\quad  \Delta_b=r^2-2(M-b)r+a^2\,.
\ee
The 3-form $H$ reads
\be\label{HHH}
H=-\frac{2\,b\,a}{\rho_b^4}\,\mathrm{d}t\wedge \mathrm{d}\phi\wedge\Bigl[\bigl(r^2-a^2\cos^2\!\theta\bigr)\sin^2\!\theta\,\mathrm{d}r-r\Delta_b\sin 2\theta \mathrm{d}\theta\Bigr]\,.
\ee
Note that, the transformation $g_{ab}\to e^{\Phi}g_{ab}$ can be performed to go from the string frame
to the Einstein frame. Our choice for the string frame is guided by the fact that, in the context of separability, the string frame seems to be more fundamental than the Einstein one, as are the hidden symmetries present in the Kerr--Sen spacetime, see \cite{Houri:2010fr}.

As mentioned in the introduction, the solution describes a black hole with mass $M$, $U(1)$ charge~\cite{Sen:1992ua}~\footnote{That this is the conserved charge of the system follows from the equation of motion \eqref{eq:HourieetalEq4.2} for $F$. See for instance~\cite{Jiang:2019vww} for an explict calculation of the charges in the Kerr--Sen spacetime.}
\be
Q=\frac{1}{4\pi}\int_{S^2_\infty}e^{\Phi}\star(F-A\cdot H)\,,
\ee
angular momentum $J=Ma$, and magnetic dipole moment $\mu_g=Qa$. When  the `twist parameter'
 \be
 b=\frac{Q^2}{2M}
 \ee
is set to zero, the solution reduces to the Kerr geometry. The horizon of the Kerr--Sen black hole is located at $r=r_+\equiv {M-b+\sqrt{(M-b)^2-a^2}}$ when the inequality $M - b \geq |a|$ holds. As in the Kerr--Newmann case  the ergosphere is  present and responsible for the instability modes but it is now located at $r_+<r<r_e\equiv {M-b+\sqrt{(M-b)^2-a^2\cos^2\theta}}$.
Moreover, the Kerr-Sen black hole also obeys the first law, \eqref{first}, where now the (Einstein frame) thermodynamic quantities are given by
\ba\label{TDsSen}
\Omega_H&=&\frac{a}{r_+^2+2br_++a^2}\;,\quad \phi_H=\frac{Qr_+}{r_+^2+2br_++a^2}\,,\nonumber\\
T_H&=&\frac{r_+^2-a^2}{4\pi r_+(r_+^2+2br_++a^2)}\,,\quad
S=\pi(r_+^2+2br_++a^2)\,.
\ea

The spacetime no longer possesses the hidden symmetry of the principal tensor. However, as shown in \cite{Houri:2019lnu} a weaker structure of the {\em principal tensor with torsion}
exists \cite{Kubiznak:2009qi}. This is a non-degenerate closed conformal Killing--Yano tensor with torsion, obeying the following generalization of Eq. \eqref{PT}:
 \be\label{PTT}
\nabla^T_{c} h_{ab}=g_{ca}\xi_b-g_{cb}\xi_a\,,\quad \xi^a=\frac{1}{3}\nabla^T_c h^{ca}\,.
\ee
Here, the covariant derivative with torsion is defined as
\ba
\nabla^T_a M^{b\dots}{}_{c\dots}&=&\nabla_a M^{b\dots}{}_{c\dots} +\frac{1}{2}T^b{}_{ad} M^{d\dots}_{c\dots}+\dots\nonumber -\frac{1}{2}T^d{}_{ac}M^{b\dots}{}_{d\dots}+\dots,
\ea
and the torsion is simply identified~\cite{Houri:2019lnu} with the 3-form $H$, \eqref{HHH},
\be
T_{abc}=H_{abc}\,.
\ee
More explicitly we have
\be\label{PTTe}
h=e^{-\Phi}\Bigl[r(\mathrm{d}t-a\sin^2\!\theta \mathrm{d}\phi)\wedge \mathrm{d}r-a\cos \theta \bigl[a\,\mathrm{d}t-(r^2+2br+a^2)\mathrm{d}\phi\bigr]\wedge \mathrm{d}\cos\theta\Bigr]\,.
\ee
Despite being a weaker structure, the principal tensor with torsion still gives rise to standard Killing tensor, via \eqref{KT}. However, the isometries of the spacetime  are no longer straightforwardly generated from $h$ \cite{Houri:2019lnu}.

%%%%%%%%%%%%%%%%%%%%%%%%%%%%%%%%%%%%%%%%%%%%%%%%%%%%%%%%%%%%%%%%%%%%%%
%%%%%%%%%%%%%%%%%%%%%%%%%%%%%%%%%%%%%%%%%%%%%%%%%%%%%%%%%%%%%%%%%%%%%%%
\section{Separability of Proca equations}\label{sec:Proca}

In this section we present the form of the Proca equations in the Kerr--Newmann geometry and motivate how this changes for the Kerr--Sen case. We then outline our ansatz and resulting separated equations using the (generalized) hidden symmetries of these two spacetimes. The full details of the calculation for the Kerr--Sen spacetime can be found in appendix~\ref{sec:Separation}.

%%%%%%%%%%%%%%%%%%%%%%%%%%%%%%%%%%%%%%%%%%%%%%%%%%%%
\subsection{Proca in Kerr--Newman spacetime}

In curved spacetime and in the absence of sources, the standard Proca equation reads

\be\label{Procaeq}
\nabla_a \mathcal{F}^{ab}-m^2P^b=0\;,
\ee
where $m$ stands for the mass of the particle and the the field strength $\mathcal{F}$ is defined in terms of the massive $U(1)$ vector field $P$ in a standard way, $\mathcal{F}_{ab}=\nabla_a P_b-\nabla_bP_a$.
Due to the presence of mass term, there is no longer gauge invariance, however \eqref{Procaeq} automatically implies the ``Lorenz condition''
\be\label{Lorenz}
\nabla_{a}P^a=0\;.
\ee

The separability of the Proca equation in the Kerr--Newmann background was demonstrated in \cite{Frolov:2018ezx} (the Kerr--Newman metric is a special case of the $D=4$ canonical metric for which the separability was shown there). Let us briefly recapitulate this result. The key step is to use the LFKK ansatz~\cite{Lunin:2017,Frolov:2018ezx,KrtousEtal:2018,FrolovKrtousKubiznak:2018a} for the gauge field $P$,
\be\label{eq:Ans}
P^a=B^{ab}\nabla_b Z\;,
\ee
where $B$ is the \emph{polarization tensor} (not to be confused with the 2-form potential $\mathcal{B}$ appearing in the definition of $H$), which can be
covariantly written in terms of the metric and the principal tensor $h$, \eqref{PTe}, as
\be\label{BB}
B^{ab}(g_{bc}+i\mu h_{bc})=\delta^a_c\;,
\ee
where $\mu$ is a separation constant. The function $Z$ assumes the standard multiplicative separation form,
\be
Z=R(r)\,S(\theta)\,e^{im_\phi \phi}\,e^{-i\omega t}\,,
\ee
where $m_\phi$ and $\omega$ are the eigenvalues of $i \partial_t$ and $-i\partial_\phi$. Note that $\phi$ has period $2\pi$, and regularity of the spherical harmonics $S(\theta)\,e^{im_\phi \phi}$ requires that $m_{\phi}\in\mathbb{Z}$.

With this ansatz, the Proca equation \eqref{Procaeq} reduces to two differential equations in $r$ and $\theta$, respectively, which only couple to each other via their dependence on the Killing parameters $\{\omega, m_\phi\}$, the separation constant $\mu$, the Proca mass parameter $m$, and the black hole parameters $\{M, Q, a\}$.\footnote{In the actual separation, it is advantageous to first concentrate on the Lorenz condition \eqref{Lorenz}. This already yields the resultant separated equations, but contains an arbitrary new separation constant. This constant is then fixed in terms of the mass $m$ and the separation constant $\mu$ by using the full Proca equation \eqref{Procaeq}, we refer to refs.~\cite{Frolov:2018ezx,KrtousEtal:2018, Cayuso:2019vyh} and App.~\ref{sec:Separation} for more details.}
These equations take the explicit form
\begin{subequations}\label{ProcaKN}
	\begin{align}
	\begin{split}
	\frac{\mathrm{d}}{\mathrm{d} r}\,
	\left[\frac{\Delta}{q_r}\frac{\mathrm{d} R}{\mathrm{d} r}\right] + \left[\frac{K_r^2}{\Delta\,q_r}{+}\frac{2-q_r}{q_r^2}\frac{\sigma}{\mu}
	-\frac{m^2}{\mu^2} \right] R =0\,,
	\end{split}
	\\
	&\frac{1}{\sin\theta}\frac{\mathrm{d}}{\mathrm{d} \theta}\,
	\left[\frac{\sin \theta}{q_\theta}\frac{\mathrm{d} S}{\mathrm{d} \theta}\right]
	-\left[\frac{K_\theta^2}{q_\theta \sin^2\theta}{+}\frac{2-q_\theta}{q_\theta^2}\frac{\sigma}{\mu}-\frac{m^2}{\mu^2}\right]S=0\,,
	\end{align}
	\label{eqs:coupled}
\end{subequations}
where
\be
\begin{gathered}
K_r = a\,m_{\phi}-(a^2{+}r^2)\omega\,,\quad K_\theta = m_{\phi}-a\,\omega\,\sin^2\theta\,,
\\
q_r = 1+\mu^2r^2\,,\quad q_{\theta} = 1-\mu^2a^2\cos^2\theta\,,
\\
\sigma = a \mu ^2 \left(m_{\phi }-a \omega \right)+\omega\,.
\end{gathered}
\ee

The demonstrated separation depends crucially on the existence of the principal tensor in a number of ways. First, the separation occurs in geometrically preferred coordinates determined by the principal tensor -- coordinates $r$ and $\cos\theta$ are related to the eigenvalues of the principal tensor, e.g. \cite{FrolovKrtousKubiznak:2017review}.
Second, the principal tensor explicitly enters the separation ansatz \eqref{eq:Ans} via the polarization tensor \eqref{BB}. Third, the principal tensor gives rise to a complete set of mutually commuting operators that guarantee this separability \cite{KrtousEtal:2018, Houri:2019lnu, Houri:2019nun}. Namely, apart from the (trivial) ones connected with Killing vectors, the following two (2nd order) operators directly link to the separation ansatz:
\be
\hat g=\nabla_a( g^{ab} \nabla_b)-2i\mu V_ag^{ab}\nabla_b\,,\quad
\hat K=\nabla_a (K^{ab} \nabla_b)-2i\mu V_a K^{ab} \nabla_b\,,
\ee
where $K^{ab}$ is the Killing tensor \eqref{KT} and $V^a=\xi_b B^{ba}$, see App.~\ref{sec:Separation} for more details.

%%%%%%%%%%%%%%%%%%%%%%%%%%%%%%%%%%%%%%%%%%%%%%%%%%%%%%%%%%%%%
\subsection{Generalized Proca in Kerr--Sen spacetime}

Test fields in the Kerr--Sen background naturally pick up modifications due to the presence of background fields $\phi, A,$ and $H$, see for example \cite{Kubiznak:2009qi,Houri:2010qc,Houri:2010fr} for a modification of the Dirac equation. To motivate the generalized Proca equation,
we assume that the massive vector field $P$ couples to the
background fields $\Phi$ and $H$ in analogy to the massless Maxwell field already present in the Kerr--Sen action \eqref{action}. As in~\cite{Cayuso:2019vyh}, we also demand that the modified Proca is linear in $P$, reduces to the Proca equation in the absence of the background fields, and obeys current conservation in the presence of sources.

It follows that there are two key modifications to the Proca equation in the Kerr--Sen background. First, in the string frame the dilaton enters the field equation
\be \label{eq:Troca}
\nabla^a \left(e^\Phi \mathcal{F}^T_{\;ab}\right)-m^2e^\Phi P_b=0\;.
\ee
Second, the 3-form $H_{abc}=T_{abc}$ contributes to the field strength in a torsion-like fashion
\be\label{eq:FStPr}
\mathcal{F}^T_{\;ab}
=(d^T P)_{ab}=\nabla_a P_b-\nabla_bP_a-P^cH_{cab}\;,
\ee
where $d^T$ is the torsion generalization of the exterior derivative, $d^T=\nabla^T\wedge\ $.\footnote{Note that of the 3 possible generalizations of the `Maxwell operator' $\nabla\cdot d P$:
\be
O_1=\nabla \cdot (d^TP)\,,\quad
O_2=\nabla^T \cdot d P\,,\quad
O_3=\nabla^T \cdot (d^T P)\,,
\ee
it is only the first one which obeys the current conservation equation and (upon including the dilatonic modification) consequently appears in \eqref{eq:Troca}.}
With this definition and using the equation of motion for $H$, \eqref{eq:HourieetalEq4.2}, the Proca field equation \eqref{eq:Troca} takes the same form as the equation of motion of the Maxwell field with the addition of the standard mass term. \eqref{eq:Troca} also implies a modified ``Lorenz condition''
\be
\nabla_a (e^{\Phi} P^a)=0\,.
\ee

To separate the {\em generalized Proca equation} \eqref{eq:Troca} in the Kerr--Sen background we exploit the same machinery as for the Kerr--Newman case, with the only difference that the principal tensor \eqref{PTe} is now replaced with the principal tensor with torsion \eqref{PTTe}. Upon this, the
LFKK ansatz \eqref{eq:Ans} continues to work, see App.~\ref{sec:Separation}, and we recover the following separated equations:
\begin{subequations}\label{ProcaSen}
\begin{align}
\begin{split}
\frac{\mathrm{d}}{\mathrm{d} r}\,
  \left[\frac{\Delta_b}{q_r}\frac{\mathrm{d} R}{\mathrm{d} r}\right] + \left[\frac{K_r^2}{\Delta_b\,q_r}{+}\frac{2-q_r}{q_r^2}\frac{\sigma}{\mu}
-\frac{m^2}{\mu^2} {-}\frac{4br\omega\mu}{q_{r}^{2}} \right] R =0\,,
\end{split}
\\
&\frac{1}{\sin\theta}\frac{\mathrm{d}}{\mathrm{d} \theta}\,
   \left[\frac{\sin \theta}{q_\theta}\frac{\mathrm{d} S}{\mathrm{d} \theta}\right]
   -\left[\frac{K_\theta^2}{q_\theta \sin^2\theta}{+}\frac{2-q_\theta}{q_\theta^2}\frac{\sigma}{\mu}-\frac{m^2}{\mu^2}\right]S=0\,,
\end{align}
\end{subequations}
where
\be
\begin{gathered}
K_r = a\,m_{\phi}-(a^2{+}r^2+2rb)\omega\,,\quad K_\theta = m_{\phi}-a\,\omega\,\sin^2\theta\,,
\\
q_r = 1+\mu^2r^2\,,\quad q_{\theta} = 1-\mu^2a^2\cos^2\theta\,,
\\
\sigma = a \mu ^2 \left(m_{\phi }-a \omega \right)+\omega\,,
\end{gathered}
\ee
which are to be compared to the Proca equations in the Kerr--Newman spacetime \eqref{ProcaKN}, and upon setting $b=0$ reduce to the those in the Kerr spacetime. Note that
the angular equation in all three cases is exactly the same, while the radial one picks up some small modifications.

Similar to the Kerr--Newman case, the separability is underlain by a complete set of  mutually commuting symmetry operators, one of which is constructed from the generalized principal tensor,
\be
\hat g=e^{-\Phi}\nabla_a (e^{\Phi} g^{ab} \nabla_b)-2i\mu V_ag^{ab}\nabla_b\,,\quad
\hat K=e^{-\Phi}\nabla_a (e^{\Phi}K^{ab} \nabla_b)-2i\mu V_a K^{ab} \nabla_b\,,
\ee
see App.~\ref{sec:Separation} for more details.

Now that we have presented and separated the Proca equations for the Kerr--Newman and Kerr--Sen black holes we can turn to studying the consequences for the instabilities of the massive vector field.

%%%%%%%%%%%%%%%%%%%%%%%%%%%
\section{Unstable modes} \label{sc:NumericsResults}
%%%%%%%%%%%%%%%%%%%%%%%%%%%

%%%%%%%%%%%%%%%%%%%%%%%%%%%
\subsection{Numerics: formulation of the problem} \label{sc:numericsformulation}
%%%%%%%%%%%%%%%%%%%%%%%%%%%

Let us now present our numerical results for finding the most unstable modes of the Proca equation in the three backgrounds. We need to solve numerically the Proca coupled pair of (radial and angular) ODEs \eqref{ProcaKN} for
the Kerr--Newman black hole and  \eqref{ProcaSen} for the Kerr--Sen black hole; the results for the Kerr black hole are obtained by setting $Q=0$ in these equations.

Using $\Delta(r_+)=0$ (for Kerr--Newman) or  $\Delta_b(r_+)=0$ (for Kerr--Sen) we can replace the mass $M$ by the outer horizon radius parameter $r_+$. We then find it convenient to work with a compact radial coordinate $y\in[0,1]$ and with a new polar coordinate  $x\in[0,1]$ related to the standard coordinates $r,\theta$ of  \eqref{metricKN}  and \eqref{Senmetric} by
\begin{equation} \label{coordtransf}
r=\frac{r_+}{1-y^2}\,,\qquad \cos\theta=2x-1\,.
\end{equation}
Note that the horizon is located at $y=0$ and asymptotic infinity is at $y=1$. For numerics it is also convenient to work with the dimensionless quantities $\{\widetilde{a}=a/r_+\,,\,\widetilde{Q}=Q/r_+\,,\,\widetilde{m}=m\,r_+\,,\,\widetilde{\omega}=\omega\, r_+\,,\,\widetilde{\mu}=\mu\, r_+\}$
(although our final results will be presented in mass units, {\it i.e.} in terms of the dimensionless quantities \eqref{units} below).
%$\{J/M^2=a/M,Q/M,m M,\omega M,\mu M \}$).
In this setting, we have two unknown functions $R(y)$ and $S(x)$ whose equations have to be solved subject to appropriate physical boundary conditions\footnote{For a systematic and detailed discussion of regularity of perturbations and associated boundary conditions, we invite the reader to see the discussions in \cite{Dias:2010eu,Dias:2010maa} and in the pedagogical topical review \cite{Dias:2015nua}.}.

We are particularly interested in searching for unstable modes as these determine the signatures of the Proca fields in gravitational wave signals. These modes have frequencies whose real part is smaller than the potential barrier height set by the Proca field mass, ${\rm Re}(\omega)< m$. A Frobenius analysis at asymptotic infinity $y=1$ then indicates that unstable modes must decay as
\be\label{BC:y1}
R \big|_{y\to 1}\sim e^{-\frac{\sqrt{\widetilde{m}^2-\widetilde{\omega}^2}}{1-y^2}}(1-y^2)^{\Sigma}\,,\quad \mbox{where}\quad \Sigma\equiv i\,(1+\widetilde{a}^2+\widetilde{Q}^2)\,\frac{(\widetilde{m}^2-2\widetilde{\omega}^2)}{2\sqrt{\widetilde{m}^2-\widetilde{\omega}^2}}\,.
\ee
Here, we have already imposed a boundary condition that eliminates a solution that grows unbounded at infinity as $e^{\sqrt{\widetilde{m}^2-\widetilde{\omega}^2}/(1-y^2)}$.

At the horizon, regularity of the perturbation in ingoing Eddington--Finkelstein coordinates requires that we impose the boundary condition,
\begin{equation}\label{BC:y0}
R\big|_{y\to 0}\sim y^{-i\,\frac{\omega-m_\phi\Omega_H}{2\pi T_H}}\,,
\end{equation}
(where $\Omega_H$ and $T_H$ are the horizon angular velocity and temperature, given in see \eqref{OmegaH}, \eqref{TDsNewman} and \eqref{TDsSen}, respectively) which excludes outgoing modes, $y^{i\,(\omega-m_\phi \Omega_H)/(2\pi T_H)}$, at the horizon.

We still need to apply boundary conditions at north and south poles of the $S^2$. Here,
regularity of the perturbations requires that $m_\phi$ is quantized to be an integer. We are
interested in unstable modes which must co-rotate with the black hole because these will extract energy from the black hole. Thus we must have $m_\phi>0$. Under these conditions, regularity requires that the perturbations behave as 
\begin{equation}\label{BC:x}
S\big|_{x\to 0}\sim x^{\frac{1}{2}|m_\phi|}\,, \qquad S\big|_{x\to 1}\sim (1-x)^{\frac{1}{2}|m_\phi|}\,,
\end{equation}
which eliminates irregular modes that would diverge as $x^{-\frac{1}{2}|m_\phi|}(1-x)^{-\frac{1}{2}|m_\phi|}$.

The boundary conditions \eqref{BC:y0} and \eqref{BC:x} are straightforwardly imposed  if we define the
new functions $q_i$, $i=1,2$, as 
\be\label{def:qs}
R(y)= e^{-\frac{\sqrt{\widetilde{m}^2-\widetilde{\omega}^2}}{1-y^2}}(1-y^2)^{\Sigma}y^{-i\,\frac{\omega-m_\phi\Omega_H}{2\pi T_H}}\,q_1(y)\,,\quad
S(x)=x^{\frac{1}{2}|m_\phi|}  (1-x)^{\frac{1}{2}|m_\phi|} \,q_2(x)\,.
\ee
and search numerically for analytical functions $q_1(y)$ and $q_2(x)$.

Our pair of Proca ODEs are coupled only via the eigenvalues $\omega$ and $\mu$. But this is a non-linear eigenvalue problem for $\omega$ and $\mu$.

%%%%%%%%%%%%%%%%%%%%%%%%%%%
\subsection{Numerics: technique} \label{sc:numericstechnique}
%%%%%%%%%%%%%%%%%%%%%%%%%%%
We discretize the field equations using a pseudospectral collocation grid on Gauss--Chebyshev--Lobbato
points. The eigenvalues and respective eigenvectors are efficiently and accurately computed using  a
powerful numerical procedure developed in gravitational contexts in
\cite{Dias:2009iu,Dias:2010maa,Dias:2010eu,Dias:2010gk,Dias:2011jg,Dias:2014eua,Dias:2010ma,Dias:2011tj,Dias:2018zjg,Dias:2013sdc,Cardoso:2013pza,Dias:2015wqa,Dias:2018ynt,Dias:2018etb,Dias:2018ufh}
-- discussed in detail in \cite{Cardoso:2013pza} and in section III.C and VI.A of the topical review
\cite{Dias:2015nua} -- which employs the Newton--Raphson root-finding algorithm.
These numerical methods are very well tested in different contexts and extremely robust. In particular, they are the same that were used to compute the ultraspinning and bar-mode gravitational instabilities of rapidly spinning Myers--Perry black holes \cite{Dias:2009iu,Dias:2010maa,Dias:2010eu,Dias:2010gk,Dias:2011jg,Dias:2014eua}, the near-horizon scalar condensation and superradiant instabilities of black holes \cite{Dias:2010ma,Dias:2011tj,Dias:2018zjg}, the gravitational superradiant instability of Kerr-AdS black holes \cite{Dias:2013sdc,Cardoso:2013pza}, the electro-gravitational quasinormal modes of the Kerr--Newman black holes \cite{Dias:2015wqa} and the modes that violate strong cosmic censorship in de Sitter backgrounds \cite{Dias:2018ynt,Dias:2018etb,Dias:2018ufh}, to mention a few. All our results have the exponential convergence on the number of gridpoints expected for a code that uses pseudospectral collocation (see \cite{Dias:2015nua}). In particular, all the results that we present in the next section are accurate at least up to the 11th decimal digit.

%%%%%%%%%%%%%%%%%%%%%%%%%%%
\subsection{Discussion of the parameter space}\label{sc:parameterspace}
%%%%%%%%%%%%%%%%%%%%%%%%%%%
Our Proca-black hole system has a scaling symmetry which we use to present the dimensionless physical quantities measured in black hole mass units, namely
\be\label{units}
\{J/M^2,Q/M,m M,\omega M,\mu M \}
\ee
(black hole angular momentum, black hole charge, Proca mass, and frequency and angular eigenvalues, respectively). Recall that $J= M a$ so the dimensionless rotation $a/M$ gives also the dimensionless angular momentum of the black hole: $J/M^2=a/M$.

The linear modes scan a 3-parameter phase space parametrized by the Proca mass $m M\geq 0$, the
black hole angular momentum $0 \leq J/M^2\leq 1$, and the electric charge $0\leq Q/M\leq
Q/M\big|_{\text{ext}}$. For Kerr--Newman extremality ({\it i.e.} zero temperature) for a given $a/M$
is attained at $Q/M\big|_{\text{ext}}=\sqrt{1-J^2/M^4}$ while the extremal Kerr--Sen black hole has
$Q/M\big|_{\text{ext}}=\sqrt{2}\sqrt{1-J/M^2}$.

In these conditions one could fix, for example, the rotation of the black hole and plot the frequency
and angular eigenvalues as a function of $Q/M$ and $mM$. Alternatively, one could fix the black hole
charge $Q/M$ or the Proca mass $mM$ and generate the corresponding 3-dimensional plots. However, we
find that these 3-dimensional plots are not particularly enlightening (especially because we want to
compare Kerr--Newman with Kerr--Sen and thus we would have two 2-dimensional surfaces on the same plot).
Therefore, we opt to produce 2-dimensional plots that are clear and describe the typical qualitative
behaviour of the system. Concretely, we will fix $J/M^2=a/M$ and $mM$ and show how the frequency
$\omega M$ and angular eigenvalue $\mu M$ change as $0\leq Q/M\leq Q/M\big|_{\text{ext}}$. The qualitative features for other values of $J/M^2$ and $m M$ are similar.

To localize ourselves in the parameter phase space it is a good idea to consider first a Proca field in the Kerr black hole background. Recall that both the Kerr--Newman and the Kerr--Sen solutions reduce to Kerr when $Q/M=0$. The properties of this solution were already studied in \cite{Cardoso:2018tly,Frolov:2018ezx,Dolan:2018dqv} and, as a test of our numerical codes for Kerr--Newman and Kerr--Sen when $Q=0$,  we have reproduced  them. As described in detail in \cite{Cardoso:2018tly,Frolov:2018ezx,Dolan:2018dqv},  Proca fields have three sectors of perturbations. Here, we will only discuss the most interesting family, namely the one that is the most unstable. For the same reason, we also only consider its lowest radial overtone case and %\sout{we only consider}
modes with azimuthal number $m_\phi=1$.

\begin{figure*}[th]
\centerline{
\includegraphics[width=.5\textwidth]{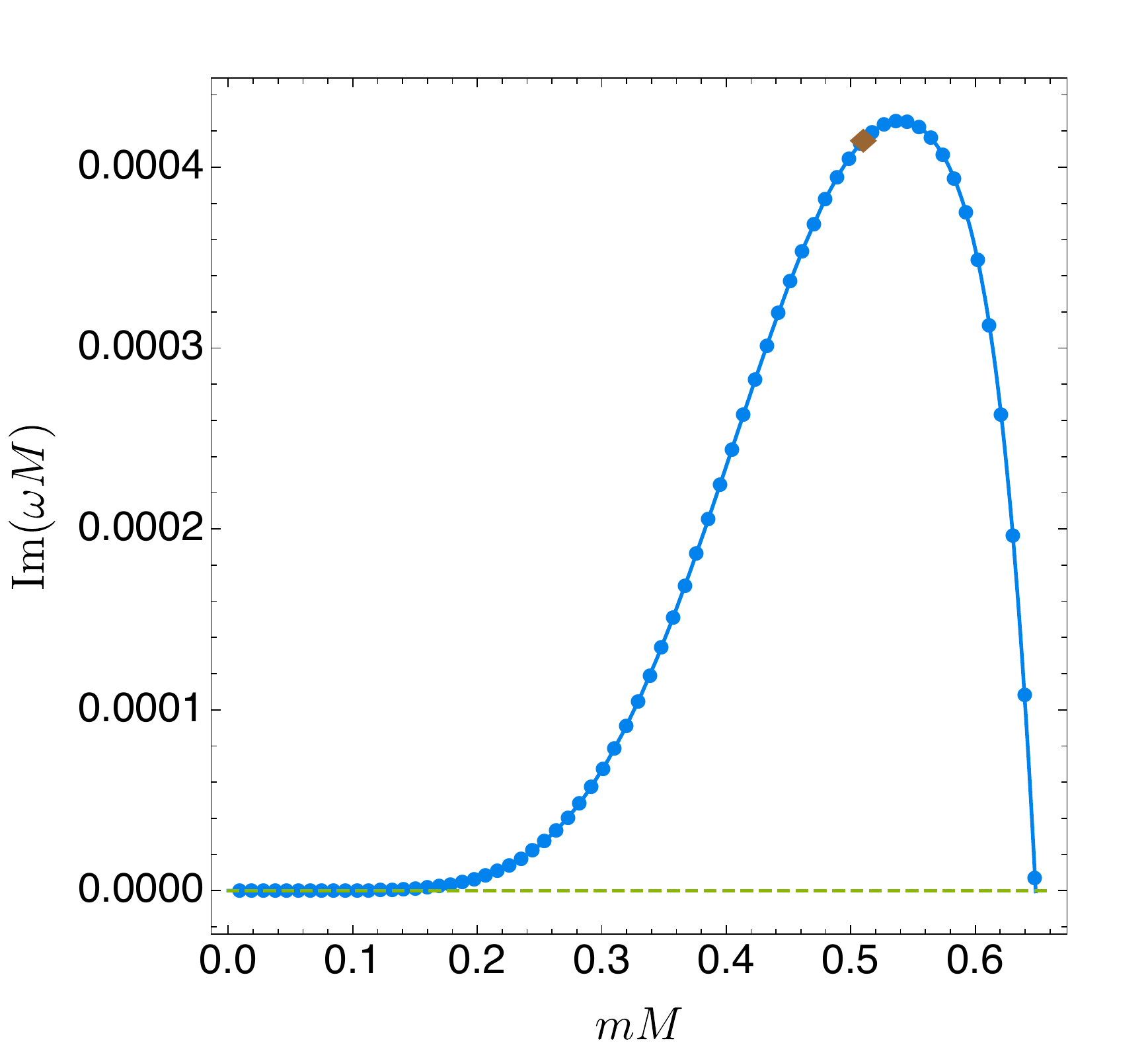}
\hspace{0.3cm}
\includegraphics[width=.425\textwidth]{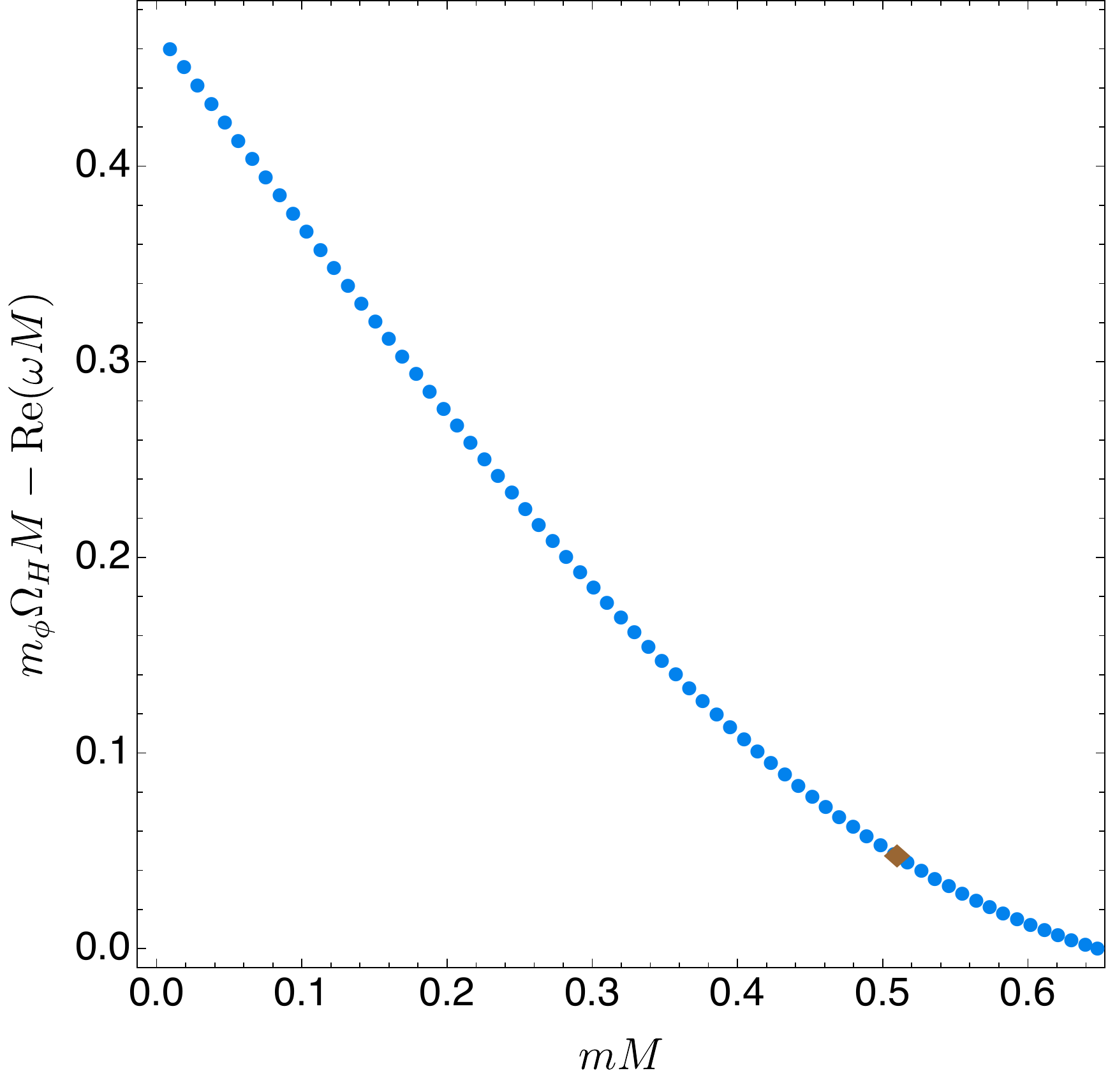}
}
\centerline{
\includegraphics[width=.5\textwidth]{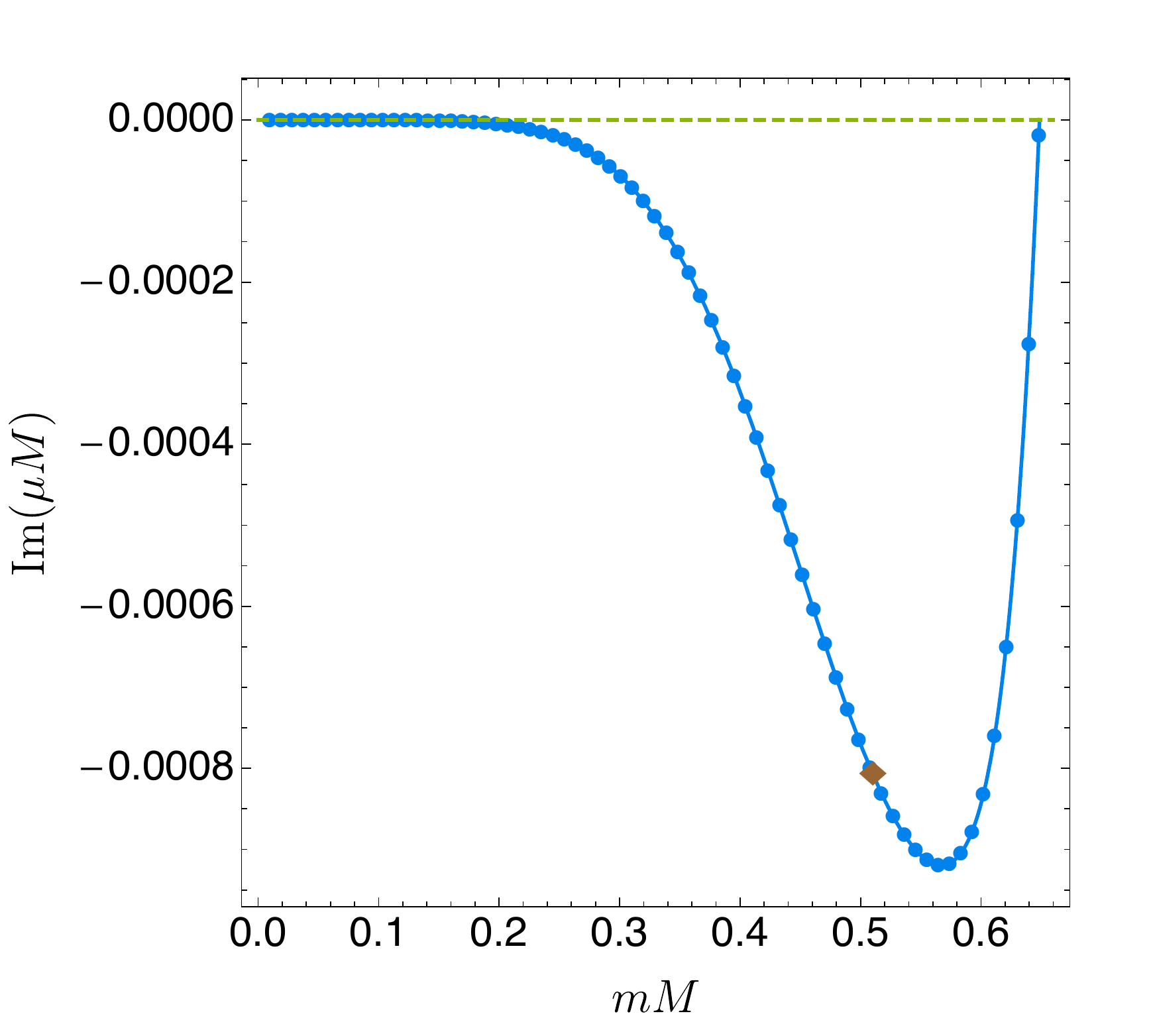}
\hspace{0.3cm}
\includegraphics[width=.47\textwidth]{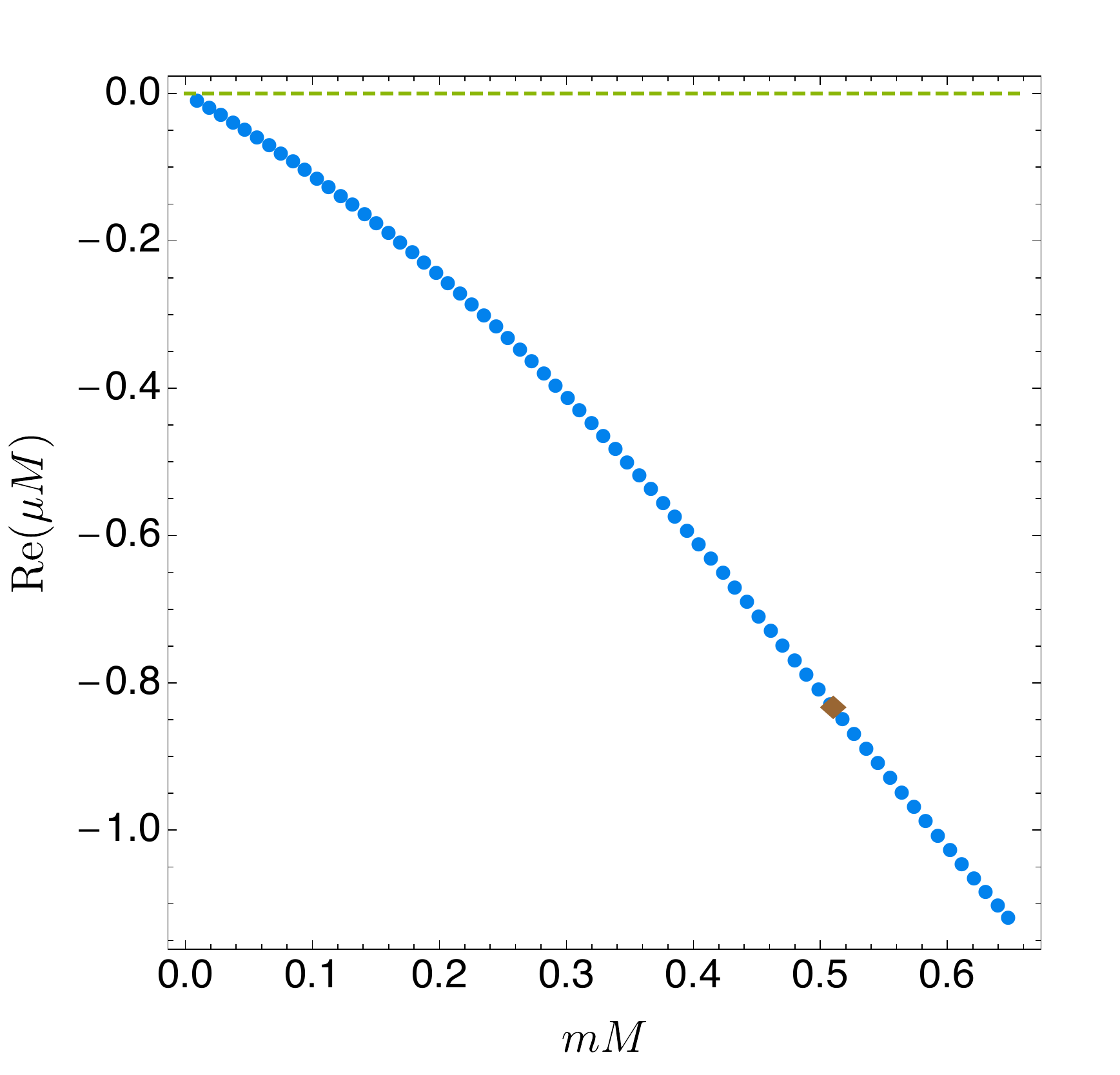}
}
\caption{{\bf Kerr black hole: effect of Proca mass.} Unstable Proca modes for a Kerr black hole with $J/M^2=a/M=0.998$ ($m_\phi=1$) as a function of the dimensionless Proca mass $mM$. The upper panels describe the imaginary (left) and real part (right) of the dimensionless frequency $\omega M$, while the lower panels give the imaginary (left) and real  (right) parts of the dimensionless angular eigenvalue $\mu M$. For this $J/M^2$ the instability is present for $0<mM\lesssim 0.64853159519$, and the maximum of ${\rm Im}(\omega M)$ occurs for $m M\sim 0.5391620000$. The brown diamond signals the solution with Proca mass $m M=0.51$ that will be analyzed in the Kerr--Newman and Kerr--Sen backgrounds (see Fig. \ref{fig:KNsen}).}
\label{fig:kerraM0998}
\end{figure*}

%%%%%%%%%%%%%%%%%%%%%%%%%%%
\subsection{Kerr: unstable modes at fixed black hole rotation}\label{sc:Kerrfixedrotation}
%%%%%%%%%%%%%%%%%%%%%%%%%%%
Firstly, in Fig. \ref{fig:kerraM0998} we fix the the rotation of the Kerr black hole to be
\be
J/M^2=a/M=0.998\,,
\ee
and we plot how the frequency $\omega M$ and angular eigenvalue $\mu M$ of the unstable modes change as we change the Proca mass $m M$. In the upper panels we give the behaviour of the dimensionless frequency $\omega M$ while in the lower panels we describe the properties of the angular eigenvalue $\mu M$. On the left-upper  panel, we analyze the imaginary part of the frequency, ${\rm Im}(\omega M)$. We see that starting from zero at $m M=0$ it first increases until it reaches a maximum at $m M\sim 0.5391620000$. For reference, at this extremum one has
\begin{eqnarray}\label{max:aM0998}
\omega M&\simeq& 0.43493600153 +0.00042563345599 \,i\,,\nonumber\\
    \mu M &\simeq& -0.89562026456 -0.00088831409834 \,i\,.
\end{eqnarray}
Then, increasing $m M$, ${\rm Im}(\omega M)$ quickly drops to zero at $mM\simeq 0.64853159519$.
The right-upper plot studies the behaviour of the real part of the frequency. More concretely, we take the difference $m_\phi \Omega_H M - {\rm Re}(\omega M)$, with $m_\phi=1$,  because the instability of the system is sourced by superradiance and this quantity should go to zero when  ${\rm Im}(\omega M)\to 0$. We find that this is indeed the case, {\it i.e.}  $m_\phi \Omega_H - {\rm Re}(\omega)=0$ at the same value $mM\simeq 0.64853159519$ where ${\rm Im}(\omega M)=0$.
Not shown in Fig.  \ref{fig:kerraM0998}, the plots for other values of fixed $J/M^2$ are qualitatively similar. The critical mass $mM$ above which the instability ceases to exist decreases when  $J/M^2$ decreases. These behaviours for the superradiant instability of massive Proca fields are similar to those found in massive scalars in the Kerr black hole, as pointed out in \cite{Cardoso:2018tly,Frolov:2018ezx,Dolan:2018dqv}. The plots in the lower panel of Fig. \ref{fig:kerraM0998}  for the angular eigenvalue speak for themselves and need few  further explanations. We just highlight that the minimum of ${\rm Im}(\mu M)$ $-$ see \eqref{max:aM0998} $-$ occurs at the same mass $m M\sim 0.5391620000$ where ${\rm Im}(\omega M)$ has its maximum and ${\rm Im}(\mu M)$ vanishes at the same $mM\simeq 0.64853159519$ as ${\rm Im}(\omega M)$.

%%%%%%%%%%%%%%%%%%%%%%%%%%%
\subsection{Kerr: unstable modes at fixed Proca mass}\label{sc:KerrfixedProcaMass}
%%%%%%%%%%%%%%%%%%%%%%%%%%%
\begin{figure*}[th]
\centerline{
\includegraphics[width=.5\textwidth]{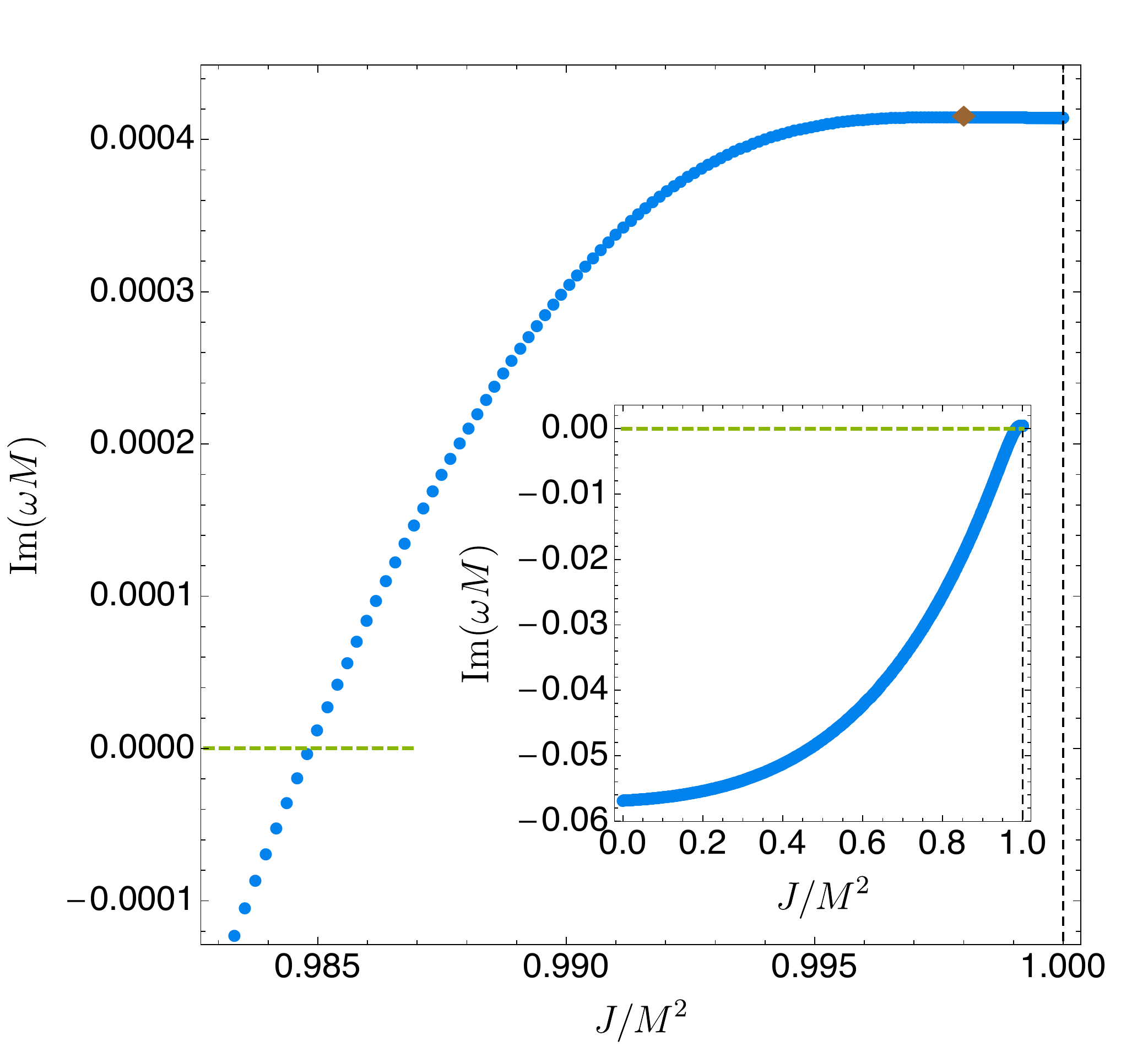}
\hspace{0.3cm}
\includegraphics[width=.46\textwidth]{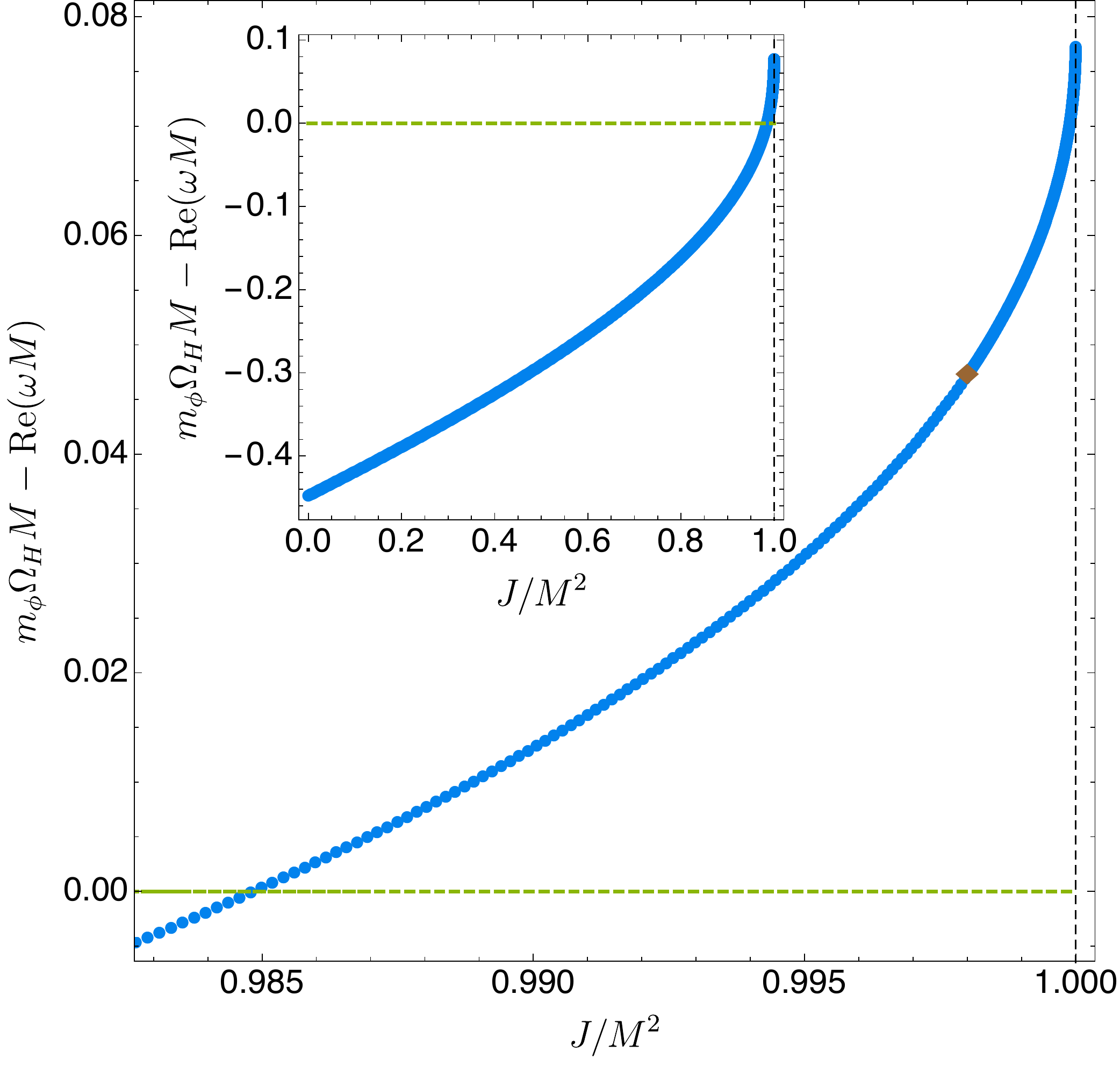}
}
\centerline{
\includegraphics[width=.5\textwidth]{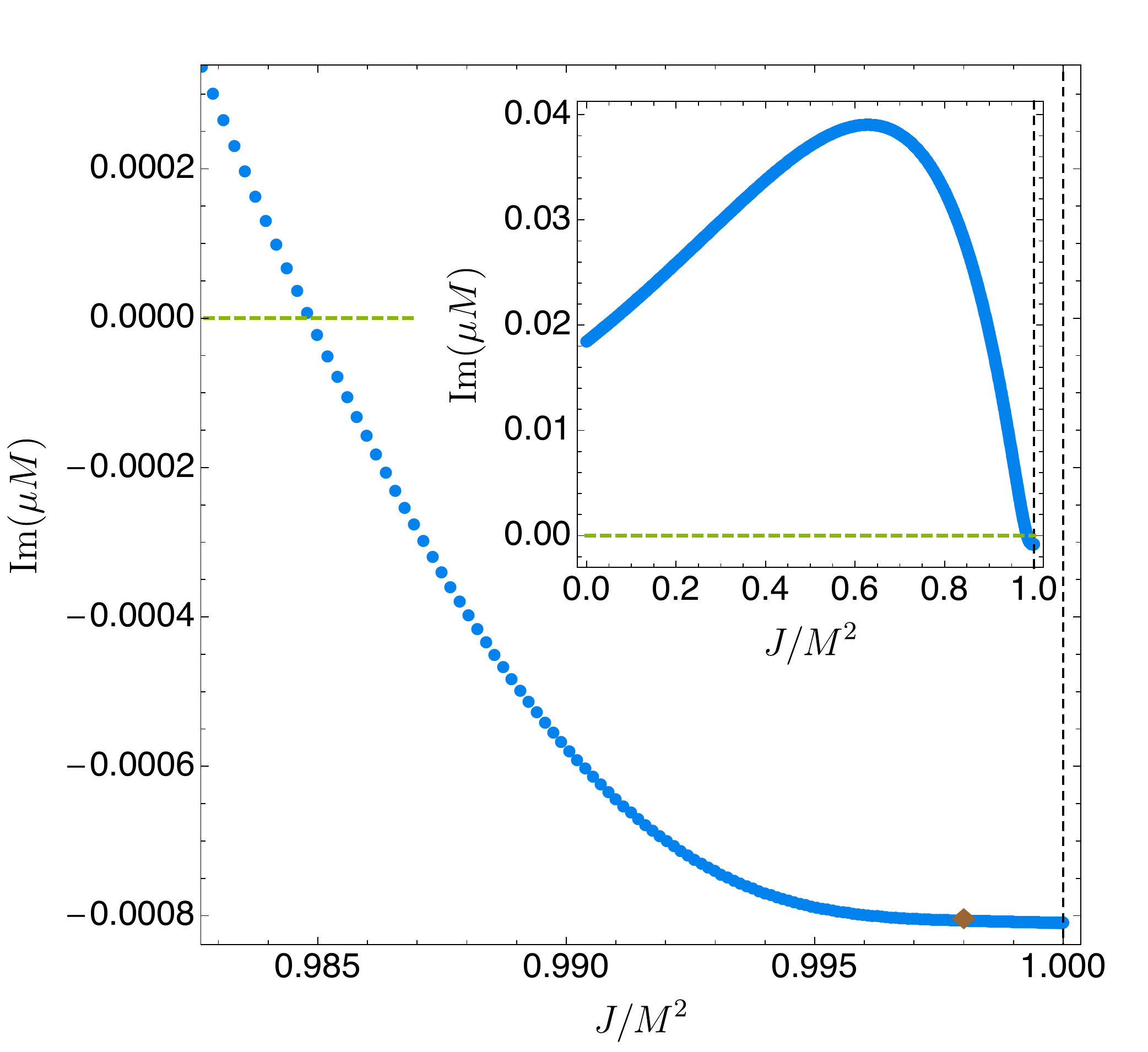}
\hspace{0.3cm}
\includegraphics[width=.49\textwidth]{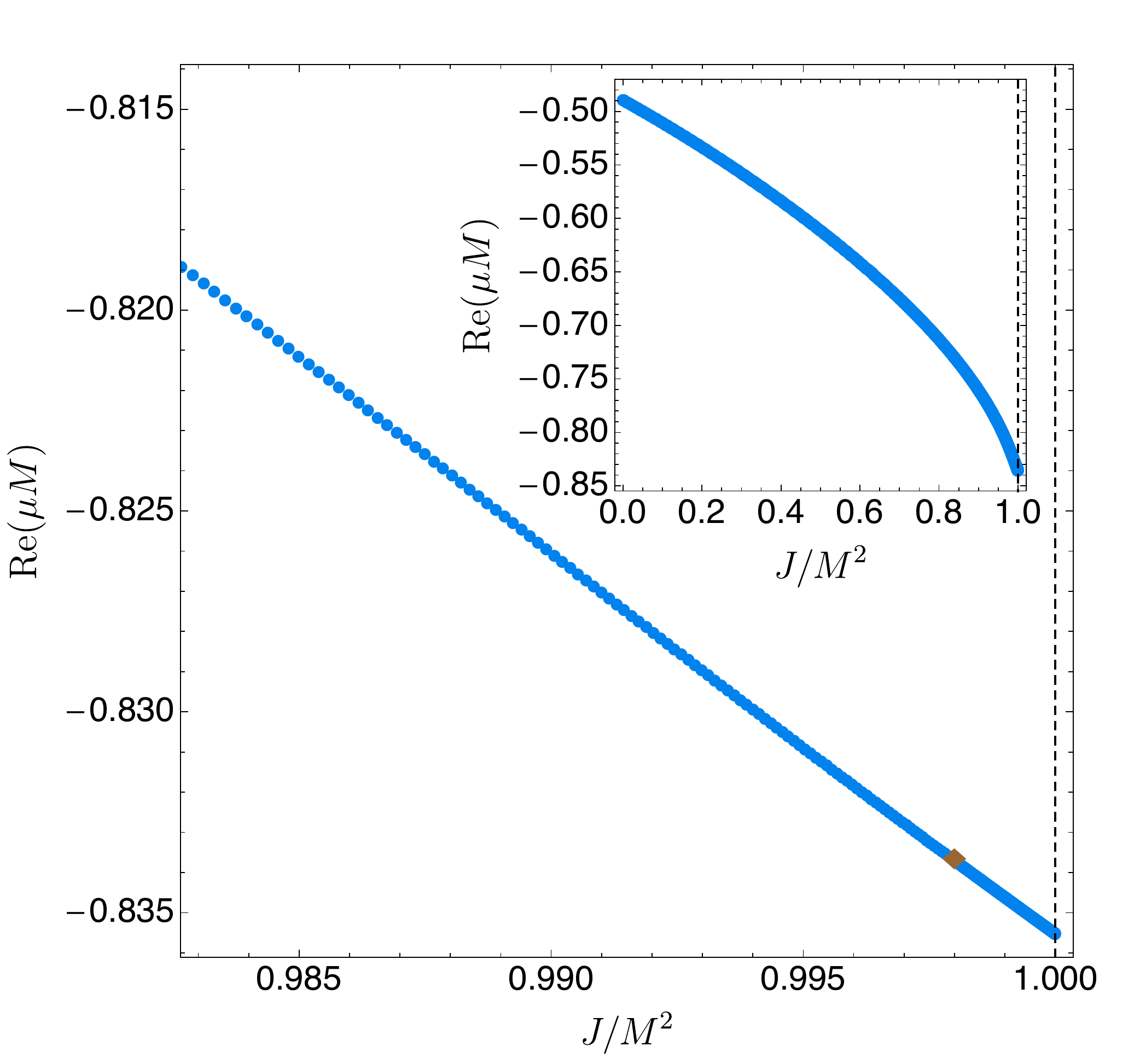}
}
\caption{{\bf Kerr black hole: effect of black hole rotation.} Unstable Proca modes with dimensionless mass $m M=0.51$ ($m_\phi=1$)  as a function of the dimensionless angular momentum $J/M^2=a/M$ of the Kerr black hole.  The upper panels describe the imaginary and real part of the dimensionless frequency $\omega M$, while the lower panels give the imaginary and real parts of the dimensionless angular eigenvalue $\mu M$. Both ${\rm Im}(\omega M)$ and ${\rm Re}(m_\phi \Omega_H-\omega)M$ change sign at $J/M^2\sim 0.98483373627$ and the main plots show the region that extends from the neighbourhood of this transition all the way to extremality at $J/M^2=1$ (the inset plots show all the range $0 \leq J/M^2 \leq 1$). The maximum of ${\rm Im}(\omega M)$ is attained for $J/M^2\sim 0.9976975410$. The brown diamond signals the solution with rotation $J/M^2=0.998$ that will be analyzed in the Kerr--Newman and Kerr--Sen backgrounds, namely in Fig. \ref{fig:KNsen} (this is the same solution that was pinpointed in Fig. \ref{fig:kerraM0998} also with a brown diamond).}
\label{fig:kerrmM051}
\end{figure*}

In the plots of Fig. \ref{fig:kerraM0998}, the brown diamond pinpoints the solution with $m M=0.51$ (and $J/M^2=a/M=0.998$). We find that this value of the Proca mass displays typical behaviour when instability modes are present and so provides a bridge to Fig. \ref{fig:kerrmM051}. Here we still consider the Kerr background but this time we fix the Proca mass to be
\be
m M=0.51\,,
\ee
and study the evolution of the frequency and angular eigenvalue as the background rotation varies from $J/M^2=0$ till the extremal $J/M^2=1$ value. In this path, the system passes again through the brown diamond point with $J/M^2=0.998$.  The sequence of plots follows the same structure of the previous figure. The main plots of Fig. \ref{fig:kerrmM051} focus on the region of large $J/M^2$ where the instability is present. To have a wider picture, the inset plots zoom out to cover the full range $0 \leq J/M^2\leq 1$. Starting from the left-upper plot, we see that $m M=0.51$ Proca modes only become unstable, {\it i.e.} ${\rm Im}(\omega M)$ turns positive, above the critical rotation value of $J/M^2 \simeq  0.9976975410$.  Further increasing the rotation, ${\rm Im}(\omega M)$ reaches a maximum at $J/M^2=0.98483373627$, where
\begin{eqnarray}\label{max:mM051}
\omega M&\simeq& 0.42207731732+0.00041468477293 \,i\,,\nonumber\\
    \mu M &\simeq& -0.83342120050 -0.00080597999349\,i\,.
\end{eqnarray}
and then it decreases monotonically towards extremality, $J/M^2=1$, where
\ba
\omega M&\simeq& 0.42243470228 + 0.00041405349126\, i\,,\nonumber\\
\mu M&\simeq& -0.83552601425-0.00080955432080\,i\,.
\ea
Again, if we fix other values of $mM$, we obtain plots that are qualitatively similar to those of Fig. \ref{fig:kerrmM051} as long as the Proca mass $mM$ is not too large, in which case there is no regime of $J/M^2$ where the system is unstable. The maximum of the instability decreases when $mM$ decreases (as shown in Fig.  \ref{fig:kerraM0998}). Altogether, the plots of Figs.  \ref{fig:kerraM0998} and \ref{fig:kerrmM051} allow to anticipate the shape of the 2-dimensional surface describing the system if we did the 3-dimensional plots $\omega M$ (or $\mu M$) as a function of $J/M^2$ and $mM$, so we do not plot this here (the reader can find these in Figs. 4 and 5 of \cite{Cardoso:2018tly}).

%%%%%%%%%%%%%%%%%%%%%%%%%%%
\subsection{Unstable modes in Kerr--Newman and Kerr--Sen black holes} \label{sc:results}
%%%%%%%%%%%%%%%%%%%%%%%%%%%
We are now ready to present our results for the unstable modes of the Kerr--Newman and Kerr--Sen black holes.
Recall again, that the brown diamond in Figs.  \ref{fig:kerraM0998} and \ref{fig:kerrmM051} (Kerr black hole) has $J/M^2=a/M=0.998$ and $mM=0.51$ for which one finds that:
\begin{eqnarray}\label{browndiamond}
\omega M&\simeq& 0.42212572826 + 0.00041466190852 \,i\,,\nonumber\\
    \mu M &\simeq& -0.83369985496 -0.00080656941824 \,i\,.
\end{eqnarray}

\begin{figure*}[th]
\centerline{
\includegraphics[width=.5\textwidth]{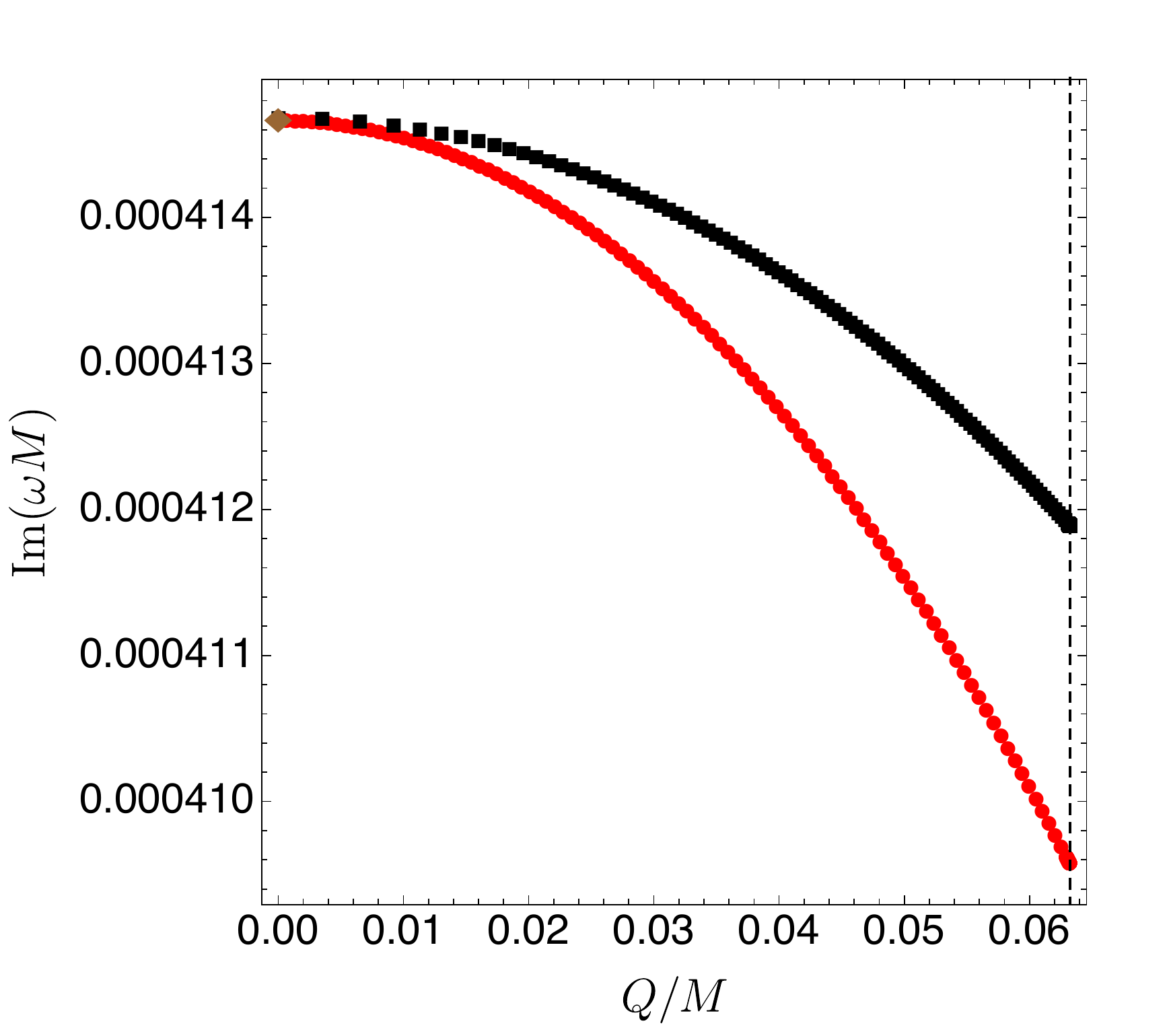}
\hspace{0.3cm}
\includegraphics[width=.47\textwidth]{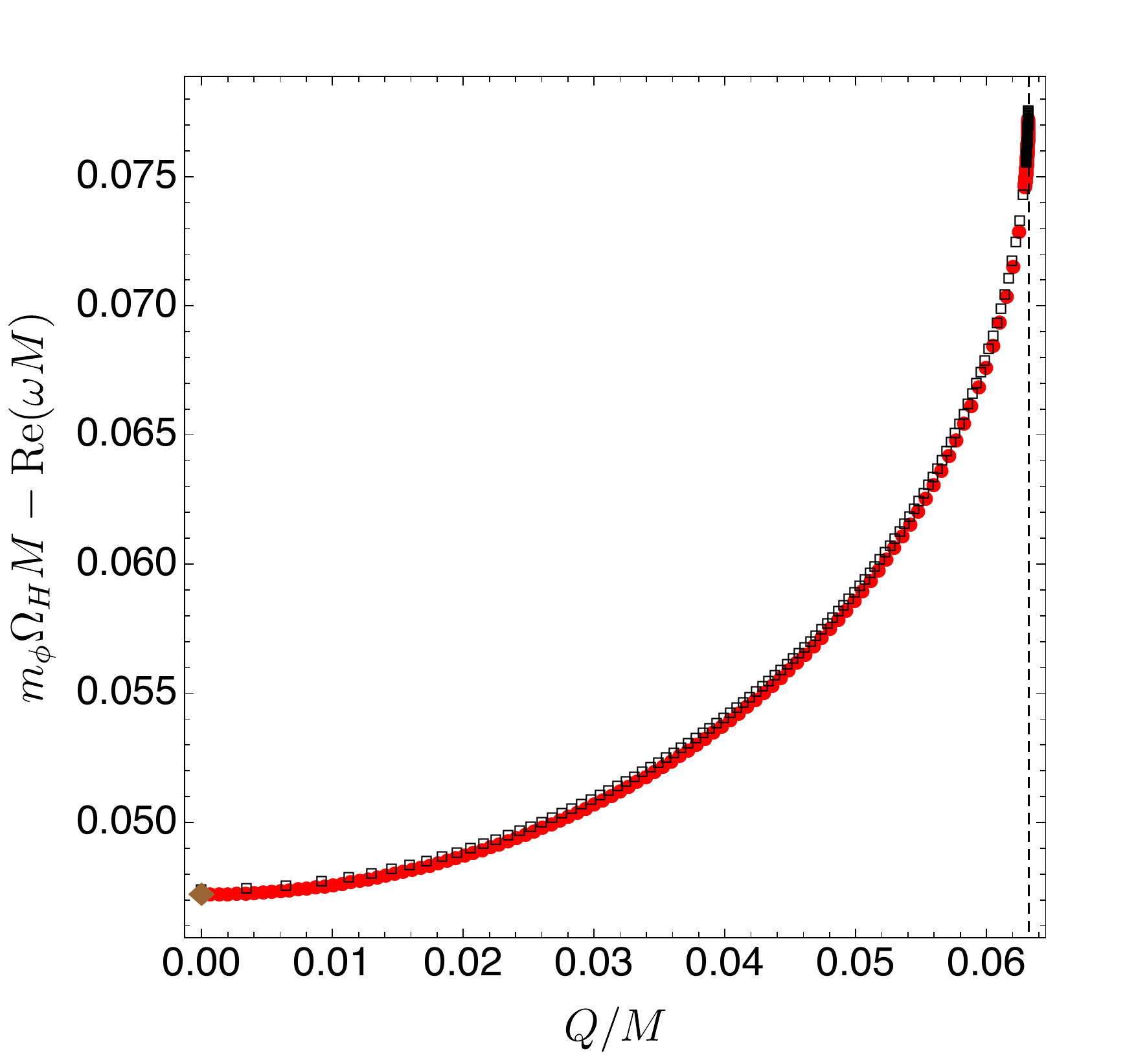}
}
\centerline{
\includegraphics[width=.5\textwidth]{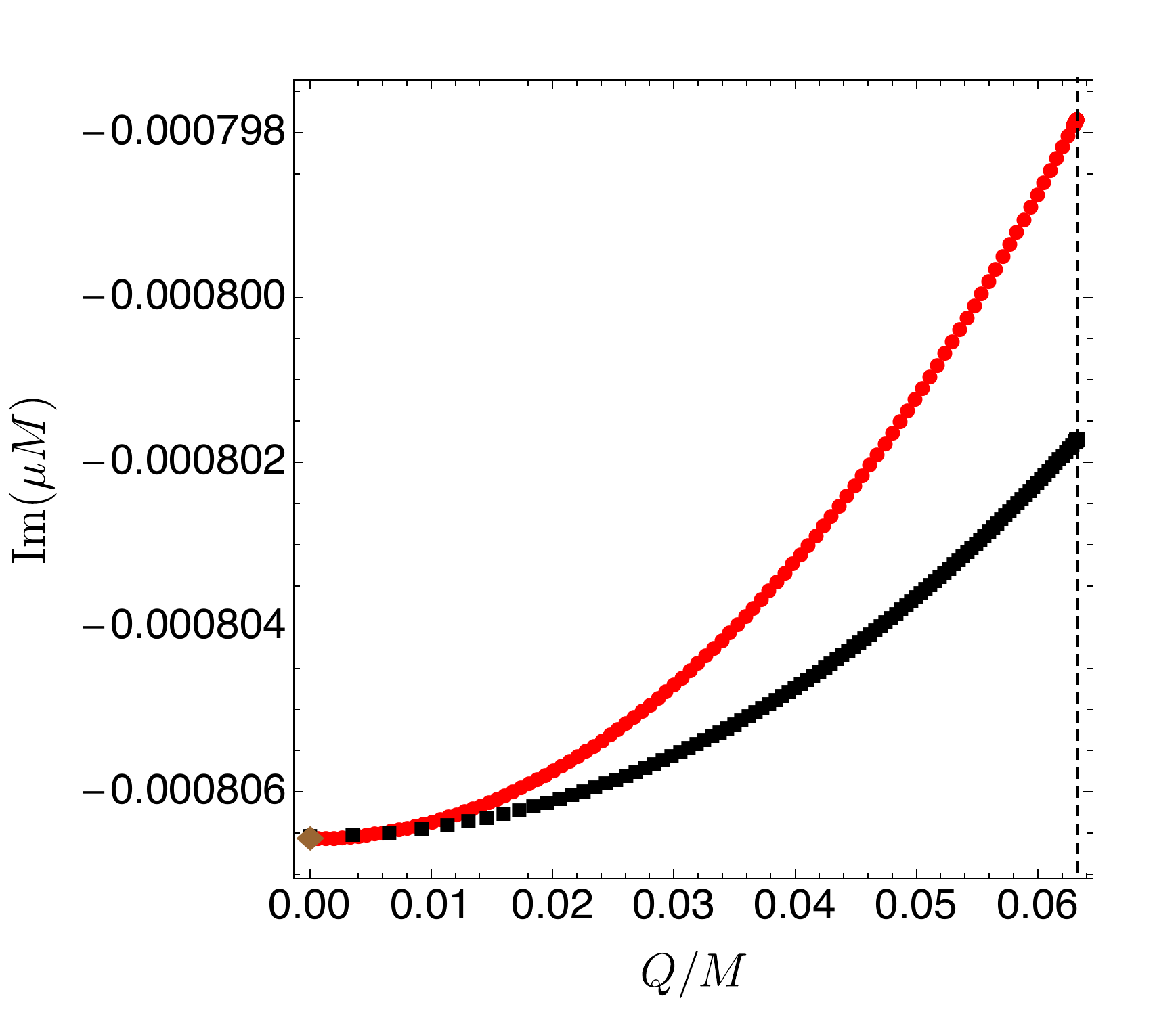}
\hspace{0.3cm}
\includegraphics[width=.47\textwidth]{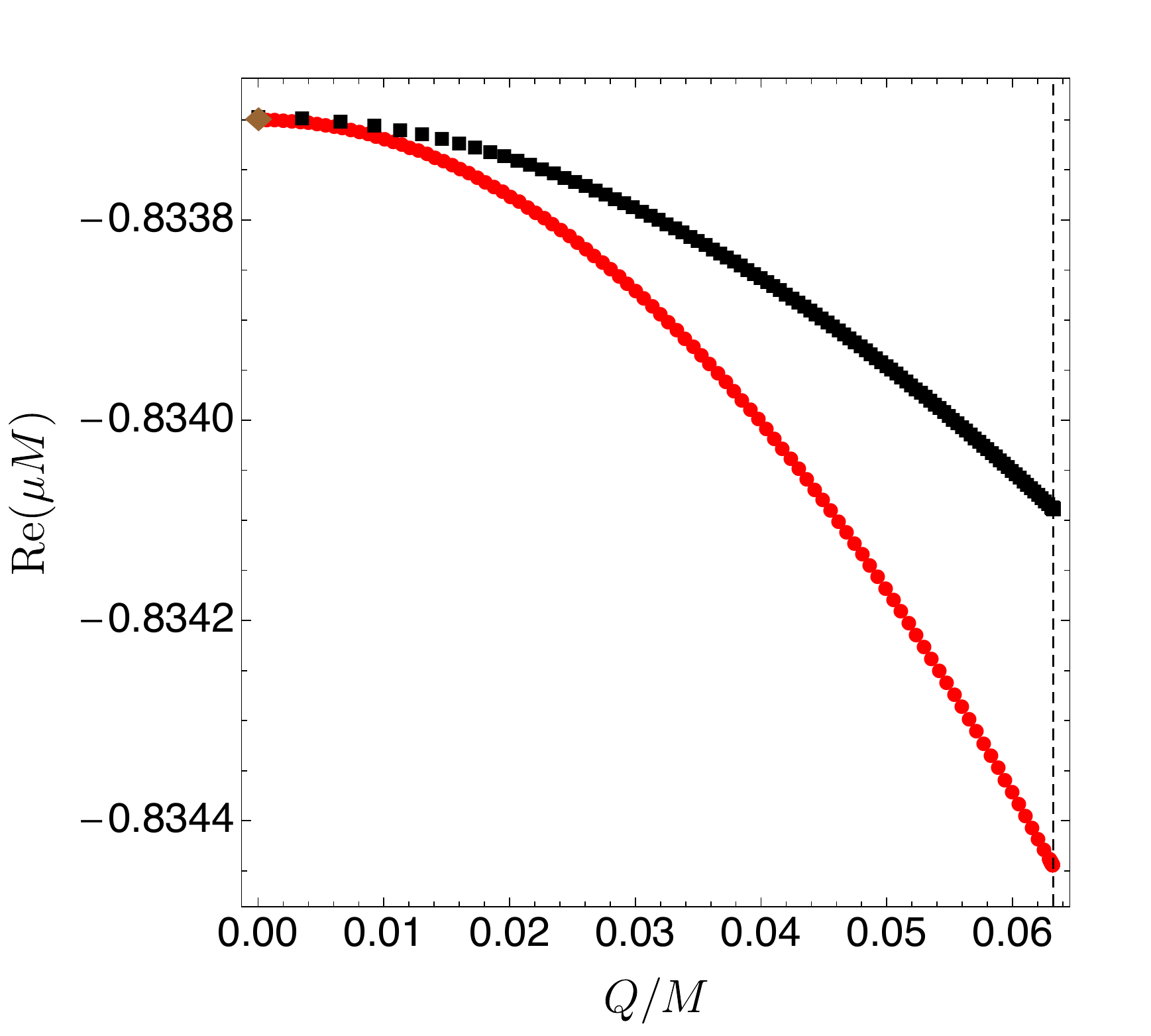}
}
\caption{{\bf Kerr--Sen vs. Kerr--Newman black holes: effect of charge.} Unstable Proca modes with mass $m M=0.51$ and $m_\phi=1$ for Kerr--Newman (red disks) and Kerr--Sen (black squares) black holes with $J/M^2=a/M=0.998$ as a function of the dimensionless hole charge $Q/M$. The upper panels describe the imaginary and real part of the dimensionless frequency $\omega M$, while the lower panels give the imaginary and real parts of the dimensionless angular eigenvalue $\mu M$. The brown diamond at $Q/M=0$ describes the Kerr solution already signalled also with a brown diamond in the plots of Figs. \ref{fig:kerraM0998} and \ref{fig:kerrmM051}.
 The vertical black dashed line signals extremality which occurs at
$Q/M=0.063213922517$ for Kerr--Newman and at $Q/M=0.063245553203$ for the Kerr--Sen black hole (so the two extremal vertical lines cannot be distinguished in the plots).}
\label{fig:KNsen}
\end{figure*}

Our strategy, to illustrate and discuss the properties of unstable Proca modes for the charged
rotating black holes, is now to take this particular Proca--Kerr solution with
$Q/M=0,J/M^2=0.998,mM=0.51$ and see how it evolves in the charged rotating black holes when the
charge increases. In Fig. \ref{fig:KNsen} we increase the charge from the Kerr limit $Q/M=0$ all the way up to the extremal limit
$Q/M=Q/M\big|_{\text{ext}}$.
As in the previous figures, the upper panels describe the imaginary (left) and real part (right) of the dimensionless frequency $\omega M$, while the lower panels give the imaginary (left) and real  (right) parts of the dimensionless angular eigenvalue $\mu M$. The red disks describe the solution in the Kerr--Newman background while the black squares describe the unstable Proca modes in the Kerr--Sen black hole.
In all the plots of Fig. \ref{fig:KNsen} the brown diamond at $Q/M=0$ is the solution \eqref{browndiamond} already pinpointed with the same diamond in the Kerr plots of  Figs.  \ref{fig:kerraM0998} and \ref{fig:kerrmM051}.  Also, the vertical black dashed line always signals extremality which occurs at $Q/M=0.063213922517$ for Kerr--Newman and at $Q/M=0.063245553203$ for the Kerr--Sen black hole (both cases are approximately given by $Q/M\sim 0.0632$ so the two extremal vertical cannot be distinguished in the plots).

The most important plot is the left-upper plot where we display the imaginary part of the frequency. As discussed above, the system is already unstable (${\rm Im}(\omega M)>0$) in the $Q/M=0$ Kerr limit (brown diamond). We then find that when the electric charge $Q/M$ is turned on, ${\rm Im}(\omega M)$ decreases monotonically $-$ both in the Kerr--Newman and Kerr--Sen black holes $-$ until it reaches a (positive) minimum at extremality: {\it the electric charge decreases the strength of the superradiant instability}. For reference, for the extremal  Kerr--Newman black hole ($Q/M=0.063213922517$) one finds
\begin{eqnarray}\label{extKN}
\omega M&\simeq& 0.42250831467 + 0.00040957576857 \,i\,,\nonumber\\
    \mu M &\simeq&-0.83444459418 -0.00079784543679 \,i\,,
\end{eqnarray}
while for the extremal Kerr--Sen black hole ($Q/M=0.063245553203$) one has
\begin{eqnarray}\label{extSen}
\omega M&\simeq& 0.42232700296 + 0.00041186820894 \,i\,,\nonumber\\
    \mu M &\simeq& -0.83409151869 -0.00080175357192 \,i\,.
\end{eqnarray}
The left-upper plot of Fig. \ref{fig:KNsen}  also shows that, in the parameter space range where both co-exist, {\it Kerr--Sen black holes are more unstable than Kerr--Newman}.\footnote{A word of caution for interpreting these results is due here regarding the use of the string vs. the Einstein frame. Namely, our calculation for the Kerr--Sen black hole has been performed in the string frame, where the Proca equations decouple and separate. It remains to be seen whether this can be directly compared to the Kerr--Newman case where the dilaton field identically vanishes.}
The other plots of Fig. \ref{fig:KNsen} are self-explanatory and need no further comments other than
that they also have a monotonic behaviour and start at the expected Proca--Kerr solution. These plots illustrate in a clear way the main properties of unstable Proca fields in Kerr--Newman and Kerr--Sen black holes. Indeed if we fix other combinations of $J/M^2$ and $mM$ (for which the instability is already present in the Kerr limit) we find similar qualitative features as those illustrated in Fig. \ref{fig:KNsen}. Thus we conclude our discussion of the most unstable Proca modes in Kerr--Newman and Kerr--Sen black holes.

%%%%%%%%%%%%%%%%%%%%%%%%%%%
\section{Conclusions}
\label{sc:conclusion}

In this paper we have shown that the LFKK ansatz can be used to separate the Proca equations in the Kerr--Sen black hole background of the low energy heterotic string theory.
This happens for a (well motivated) modification of these equations and in the string frame. It is a highly non-trivial result as the Kerr--Sen black hole no longer admits the principal tensor, which is the key object for the LFKK ansatz, and only its much weaker (torsion) generalization is present.

We have then used the resulting separated ordinary differential equations to study the corresponding instability modes of the Proca field in the Kerr--Sen background and compared them to the instability modes in the Kerr and Kerr--Newman backgrounds. This is the first study of the Proca instability modes around rotating black holes which considered a possibility of weakly charged solutions. Moreover we have considered an astrophysically viable setting where the black holes are highly spinning (close to extremal) and weakly charged. Our results allow one to compare the prediction of the two theories: the Einstein--Maxwell theory (represented by the Kerr--Newman solution) and the low energy heterotic string theory (with the corresponding Kerr--Sen black hole). Our findings indicate that, at equal asymptotic charges, Kerr-Newman black holes are more stable than Kerr--Sen ones.

Finally, our work not only generalizes the exploitation of black hole superradiance for detecting possible dark matter candidates to include charged black holes, it also opens new horizons for applications of the generalized hidden symmetries (which may not be so weak structure as previously expected).

%%%%%%%%%%%%%%%%%%%%%%%%%%%%%%
\section*{Acknowledgements}
\label{sc:acknowledgements}

Some of us (R.~C., F.~G., D.~K., A.~M., R.~G.~S, and L.~T.) would like to thank Ma{\"i}t{\'e} Dupuis and Lenka Bojdova for organizing the 2019 PSI Winter School where this project was initiated and the PSI program for facilitating this research. The work was supported in part by the Natural Sciences and Engineering Research Council of Canada. Research at Perimeter Institute is supported in part by the Government of Canada through the Department of Innovation, Science and Economic Development Canada and by the Province of Ontario through the Ministry of Economic Development, Job Creation and Trade. O.~D. acknowledges financial support from the STFC Ernest Rutherford grants ST/K005391/1 and ST/M004147/1 and from the STFC ``Particle Physics Grants Panel (PPGP) 2016" Grant No. ST/P000711/1. J.~E.~S. is supported in part by STFC grants PHY-1504541 and ST/P000681/1. J.~E.~S. also acknowledges support from a J.~Robert~Oppenheimer~Visiting~Professorship.
R.G.S thanks IFT-UNESP/ICTP-SAIFR and CNPq grant No. 132286/2018-1 for partial financial support. L.~T. acknowledges support by the Studienstiftung des Deutschen Volkes.

%%%%%%%%%%%%%%%%%%%%%%%%%%%%%%%%%%%%%%%%%%%%%%%%%%%
%%%%%%%%%%%%%%%%%%%%%%%%%%%%%%%%%%%%%%%%%%%%%%%%%%%
\appendix
\section{Separation of Proca equations in Kerr--Sen background}\label{sec:Separation}

%%%%%%%%%%%%%%%%%%%%%%%%%%%%%%%%%%%%%%%%%%%%
\subsection{Carter form of the metric}
The separation of the modified Proca equation \eqref{eq:Troca} in the Kerr--Sen background is easiest when the metric is expressed in the pseudo-Euclidean Carter-like coordinates $(\psi_0,\psi_1,x_1,x_2)$ \cite{FrolovKrtousKubiznak:2017review}. In fact, in these coordinates a more general solution to the heterotic string theory action \eqref{action}, which includes a NUT parameter,  can easily be written and reads
\cite{Houri:2010fr}:
\ba\label{generalized}
\mathrm{d}s^2&=&\frac{U_1}{X_1}\mathrm{d}x_1^2+\frac{U_2}{X_2}\mathrm{d}x_2^2+\frac{X_1}{U_1}A_1^2+\frac{X_2}{U_2}A_2^2\,,\quad e^{\Phi}=\frac{U_m}{U_1}\,,\nonumber\\
A&=&\frac{2\,c\,s}{U_m}\Bigl(m_1x_1(\mathrm{d}\psi_0+x_2^2\,\mathrm{d}\psi_1)-m_2 x_2(\mathrm{d}\psi_0+x_1^2\mathrm{d}\psi_1)\Bigr)\,,\nonumber\\
{\cal B}&=&\frac{s}{c}\bigl(\mathrm{d}\psi_0-c_0 \mathrm{d}\psi_1\bigr)\wedge A\,,
\ea
with the field strengths $F=\mathrm{d}A$ and $H=\mathrm{d}{\cal B}-A\wedge F$.\footnote{To recover the notations in the main text, one has to set
$A\to A/\sqrt{2}$. The other fields $\Phi$, ${\cal B}$, and $H$ are unchanged but the coupling between $H$ and $A$ picks up a factor of $2$ to compensate the redefinition of $A$, that is, $H\to \mathrm{d}{\cal B}-2A\wedge F=H$.}
Here
\ba
U_1&=&x_2^2-x_1^2=-U_2\,,\quad U_m=x_2^2-x_1^2-2m_1s^2x_1+2m_2s^2x_2\,,\nonumber\\
A_1&=&\frac{U_1}{U_m}\Bigl[\mathrm{d}\psi_0+\mathrm{d}\psi_1(x_2^2+2m_2 x_2s^2)\Bigr]\,,\quad
A_2=\frac{U_1}{U_m}\Bigl[\mathrm{d}\psi_0+\mathrm{d}\psi_1(x_1^2+2m_1 x_1s^2)\Bigr]\,,\quad
\ea
and the metric functions take the following form:
\be\label{Xsolutions}
X_1=c_0-2m_1 x_1+x_1^2\,,\quad X_2=c_0-2m_2 x_2 +x_2^2\,.
\ee
Here, $s=\sinh\delta$, $c=\cosh\delta$, and $c_0, m_1, m_2, \delta$ are arbitrary constants, related to the rotation parameter, mass and NUT charges, and the twist parameter.

The Kerr--Sen solution in the main text is recovered upon the following change of coordinates and parameters:
\be\label{eq:CoordTrans}
(\psi_0,\psi_1,x_1,x_2)=(t-a\,\phi,\phi/a, i\,r,a\cos\theta)\,.
\ee
Here also send $m_1s^2\rightarrow ib$, $ i m_1\rightarrow (b-M)$ , turn off the NUT parameter by setting $m_{2}=0$, and send $c_0\rightarrow -a^2$. Thence the metric functions become
\ba\label{eq:Xpoly}
	X_1 &=& 2r(M-b)-r^2-a^2\equiv -\Delta_b\,,
	\quad X_2 =-a^2\sin^2\theta\,,\\
U_1&=&r^2+a^2\!\cos^2\!\theta=\rho^2\,,\quad U_m=\rho_b^2=\rho^2+2br\,.
\ea

In what follows we shall work with a more general metric \eqref{generalized}--\eqref{Xsolutions}. In fact, as already observed for the Kerr--NUT-(A)dS metrics~\cite{Frolov:2018ezx, KrtousEtal:2018}, the separability actually works for a more general class of {\em off-shell metrics} where
\be
X_\mu=X_{\mu}(x_{\mu})
\ee
are arbitrary functions of one variable. Thence in what follows we shall leave $X_\mu(x_\mu)$ arbitrary.

The off-shell metric admits a  generalized principal tensor with torsion, which reads~\cite{Houri:2019lnu}
\be\label{eq:PrincHero}
h= x_1 \, \mathrm{d} x_{1}
\wedge {A}_1+ x_2 \, \mathrm{d} x_{2}
\wedge {A}_2\,,
\ee
and obeys the defining equation \eqref{PTT} with
\begin{equation}\label{PrimaryVec}
\xi=e^\Phi\partial_{\psi_0}\;,
\end{equation}
and the torsion identified with the 3-form $H$,
\be
T=H=\frac{s}{c}F\wedge\xi=-\left(\frac{\partial\Phi}{\partial x_1}\, \mathrm{d}x_1\wedge A_1 +\frac{\partial\Phi}{\partial x_2}\, \mathrm{d}x_2\wedge A_2 \right)\wedge\xi\,.
\ee
The associated irreducible Killing tensor is given by
\be\label{eq:KT}
K_{ab}=h_{ac}h^{c}_{\,b}-\frac{1}{2}g_{ab}h^2\;.
\ee

%%%%%%%%%%%%%%%%%%%%%%%%%%%%%%%%%%%%
\subsection{Separability of Proca equations}
\label{sc:separability}

Let us now apply the LFKK ansatz to separate the Proca equations in the generalized beckground \eqref{generalized}. As argued in Sec.~\ref{sec:Proca} the Proca equation in this background reads
\be\label{Proca}
\nabla_{\!n}\left(e^{\Phi}\procaF^{na}\right) -m^2e^{\Phi} P^a=0\,,
\ee
and implies the corresponding Lorenz condition
\be\label{Lor}
\nabla_{\!a}\left(e^{\Phi}\procaA^a\right)=0\, .
\ee

In order to separate these equations, we employ the LFKK ansatz \cite{Lunin:2017,FrolovKrtousKubiznak:2018a,KrtousEtal:2018,Frolov:2018ezx},
\be\label{ans}
\procaA^a= B^{ab} \nabla_{\!b}Z\, ,\quad
B^{ab} (g_{bc}+i\mu h_{bc}) = \delta^a_c\,,
\ee
where $\mu$ is a complex parameter, $h_{bc}$ in the generalized principal tensor \eqref{eq:PrincHero}, and the potential function $Z$ is written in the multiplicative separated form
\be\label{multsep}
Z=R_1(x_1)R_2(x_2)e^{iL_0\psi_0}e^{iL_1\psi_1}\, .
\ee

Similar to ref.~\cite{KrtousEtal:2018,Frolov:2018ezx} we first concentrate on the Lorenz condition \eqref{Lor}, for which the ansatz \eqref{ans} yields:
\be
\nabla_{\!a}\left(e^{\Phi}\procaA^a\right) =e^{\Phi} \frac{Z}{q_1q_2}\,\left( \frac{q_2}{U_1}\,
\frac{1}{R_1(x_1)}{\cal D}_1 {R}_1(x_1)+\frac{q_1}{U_2}\,
\frac{1}{R_2(x_2)}{\cal D}_2 {R}_2(x_2)\right)\, ,\label{AA}
\ee
where the differential operators are given by
\begin{align}\label{eq:Derivs}
\mathcal{D}_\mu&=q_{\mu}\frac{\pa}{\pa x_\mu}\biggl[\frac{X_{\mu}}{q_{\mu}}\frac{\pa}{\pa x_\mu}\biggr]
- \frac1{X_\mu} \Bigr[(-x_\mu^2-2m_\mu s^2 x_\mu)L_0+L_1\Bigr]^2\nonumber\\
&\quad-\frac{2-q_{\mu}}{\mu q_{\mu}}\Bigr[L_0+(-\mu^2)L_1\Bigr]-\frac{4\mu L_0\, m_\mu s^2 x_{\mu}}{q_\mu}\, ,
\end{align}
and
\be
q_{\nu}=1-\mu^2 x_{\nu}^2\,.
\ee
The Lorenz condition \eqref{Lor} will be satisfied provided the mode functions $R_{\nu}$ obey the separated equations
\be\label{DD}
{\cal D}_\nu R_{\nu}=(\scs_1-x_\nu^2\scs_0) R_{\nu}\,.
\ee
Here $C_0$ and $C_1$ are two new separation constants. Then expression in \eqref{AA} reduces to
\begin{equation}\label{AAA}
\nabla_{\!a}(e^{\Phi}\procaA^a) = e^{\Phi}\frac{Z}{q_1q_2}\,\left[ \scs_0+\scs_1\,(-\mu^2) \right]\;,
\end{equation}
and we see that the Lorenz condition holds provided we fix
\begin{equation}\label{C=0}
\scs_1=\frac{\scs_0}{\mu^2}\;.
\end{equation}
At this stage we are left with one new separation constant $C_0$ but this will be fixed by solving the full Proca equations.

The results of \cite{KrtousEtal:2018} can be also used to find the representation of the Proca equation~(\ref{Proca}) for the ansatz \eqref{ans}. Employing the Lorenz condition \eqref{Lor} one finds
\be
\nabla_{\!n}\left(e^{\Phi}\procaF^{na}\right) -m^2 e^{\Phi}\procaA^a=  e^{\Phi} B^{am}\nabla_{\!m} J\, .
\ee
Here we have introduced the object
\be\label{eq:J}
J=e^{-\Phi}\nabla_a(e^{\Phi}g^{ab}\nabla_bZ)-2i\mu\xi_aB^{ab}\nabla_bZ-m^2Z\;.
\ee
At this stage, by employing the LFFK ansatz and enforcing the Lorenz condition, the Proca equation has been reduced to solving a scalar ``wave equation''.

In particular this ``wave equation'' may be written as an eigenvalue problem
\be
\hat g Z
=m^2Z\,,
\ee
where
\be
\hat g=e^{-\Phi}\nabla_a(e^{\Phi}g^{ab}\nabla_b)-2i\mu V_a g^{ab}\nabla_b\,,\quad
V^a=\xi_bB^{ba}\,.
\ee
 In this suggestive form, where we consider the metric tensor as the trivial Killing tensor $K^{(0)}_{ab}=g_{ab}$, one can guess from the Kerr--Newman case that this operator can be generalized to the two commuting operators in 4 dimensions which are enough to guarantee the separability of this equation. Thus we define
\be
\hat K=e^{-\Phi}\nabla_a(e^{\Phi}K^{ab}\nabla_b)-2i\mu V_a K^{ab}\nabla_b\,,
\ee
where $K_{ab}$ is the Killing tensor generated from the principle tensor \eqref{eq:KT}. Then one can explicitly check that these two operators commute, i.e.
\be
\left[\hat g, \hat K\right]=0\;,
\ee
and that the solution $Z$ is also an eigenvector of $\hat K$
\be
\hat K Z=\left(\frac{m^2}{\mu^2}+\frac{L_0}{\mu}+\mu L_1\right)Z\;.
\ee
These operators are just a torsion generalization of those presented in~\cite{KrtousEtal:2018, Houri:2019lnu, Houri:2019nun} and we expect that this construction can be generalized to all dimensions.

Thus the separability of $J$~\eqref{eq:J} is guaranteed and in fact $J$ separates in the form
\be{
	J= Z \sum_{\nu=1}^\dg \frac{1}{U_\nu}\frac1{R_\nu}
	\bigl[\mathcal{D}_{\nu}-m^2 (-x_{\nu}^2)\bigr]R_\nu\, ,}
\ee
where $D_\mu$ is same the operator defined in \eqref{eq:Derivs}. In the above expression we have used the identity,
\be
\sum_{\nu=1}^\dg \frac{1}{U_{\nu}}\,(-x_{\nu}^2)^{1-j}=\delta^j_0
\ee
for $j=0$, to rewrite the mass term. This identity further ensures ${J=0}$, provided the modes $R_{\nu}(x_{\nu})$ obey separated equations \eqref{DD} and additionally the extra free separation constant $C_0$ is given by the Proca mass,
\begin{equation}\label{C0m2}
{\scs_0=m^2\;.}
\end{equation}
Summarizing, the Proca equation \eqref{Proca} for the vector field $\procaA$ in the off-shell Kerr--Sen--NUT background \eqref{generalized} can be solved by using the LFKK ansatz  \eqref{ans}, \eqref{multsep}, where the mode functions $R_\nu$ satisfy ODEs \eqref{DD}. Moreover the separation constant $\scs_0$ is given by \eqref{C0m2} and $\mu$ satisfies \eqref{C=0}. To translate our separation \eqref{DD} back into the Boyer--Lindquist form presented in the main text, \eqref{ProcaSen}, we perform the map outlined above in \eqref{eq:CoordTrans},~\eqref{eq:Xpoly}. Furthermore, we need to modify the eigenvalues $L_0$ and $L_1$, to the eigenvalues of $i \partial_t$ and $-i\partial_\phi$, $\omega$ and $m_\phi$. This is simply done via the linear map
\be
L_0 = -\omega\,,\qquad L_1 = a(m_\phi-a \omega)\,.
\ee

\bibliographystyle{JHEP}
\bibliography{references}

\end{document}